# Insights on the volume variation of $LiCo_xMn_{2-x}O_4$ electrode upon cycling: a correlative *operando* XRD and XAS study

Eunike Mahayoni[1], Louiza Larbi[2], Romain Batiot[2], Anatole Weill[2], Damien Brégiroux[2], Antonella Iadecola[1,3,*], Arnaud Perez[2,3,*]

[1] *Sorbonne Université, CNRS, Physicochimie des Electrolytes et Nanosystèmes interfaciaux (PHENIX), F-75005, Paris, France*

[2] *Sorbonne Université, CNRS, Laboratoire de Chimie de la Matière Condensée de Paris (LCMCP), F-75005, Paris, France*

[3] *Réseau sur le Stockage Electrochimique de l'Energie (RS2E), FR CNRS 3459, F-80039, Amiens, France*

[*] Corresponding author: arnaud.perez@sorbonne-universite.fr, antonella.iadecola@sorbonne-universite.fr

## Abstract

Understanding volume changes in high-voltage spinel cathodes is essential for mitigating mechanical degradation and improving cycling stability. However, linking these changes to cation-specific redox processes remains challenging. Here, we combine *operando* X-ray diffraction (XRD) and X-ray absorption spectroscopy (XAS) to investigate the $LiCo_xMn_{2-x}O_4$ spinel system, which exhibits a cubic structure and isotropic lattice response, in order to correlate the evolution of the local and long-range structure. Five compositions (x = 0, 0.25, 0.5, 0.75, and 1.0) were synthesized and electrochemically characterized, with x = 0, 0.5, and 1.0 examined in detail under identical cycling conditions. XAS at the Mn and Co K-edges reveals distinct $Mn^{4+}/Mn^{3+}$ and $Co^{4+}/Co^{3+}$ redox processes, with Mn redox activity dominating below ~4.7 V and Co redox triggered at higher potentials. Chemometric tools for XAS data analysis further resolve the evolution of local structure during cycling. Both XRD and XAS capture consistent trends in volume evolution, despite probing different structural length scales. Increasing Co content progressively decreases the overall volume change, from 5.0% to 1.6% by XAS and from 5.8% to 2.1% by XRD. These results demonstrate that Co substitution suppresses electrochemically induced structural strain and establish a direct link between cation-specific redox activity and volumetric evolution. The correlative *operando* XRD/XAS approach provides a framework to investigate newly designed low-strain, high-energy cathodes with improved cycling stability.

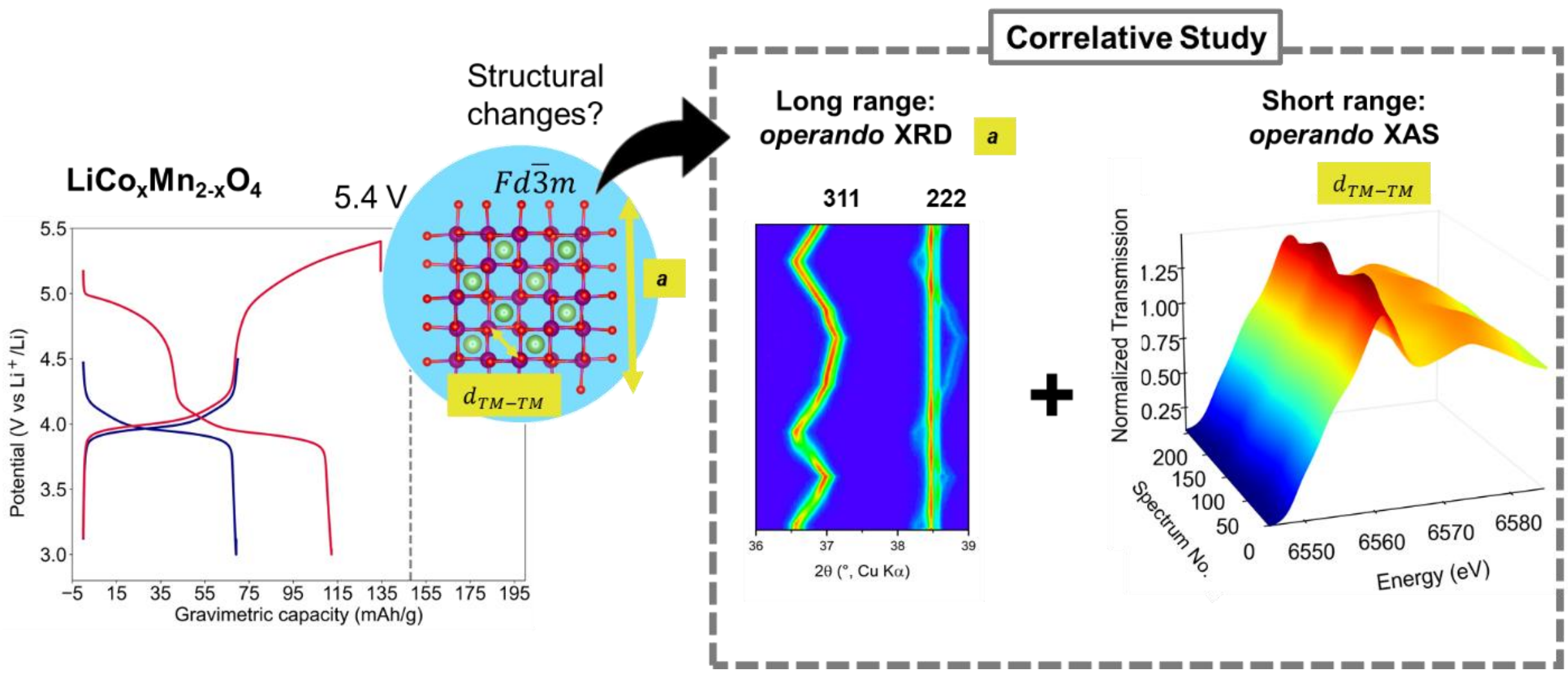

## 1. Introduction

The selection of electrode materials for rechargeable Li-ion batteries considers several performance criteria such as volumetric and gravimetric energy density, capacity retention, charging rate, lifespan and safety. The shift of paradigm from conventional Li-ion to the so-called “all solid-state battery (ASSB)” technology, which consists in replacing the liquid organic electrolyte by a solid electrolyte, forces material chemists to redefine these selection criteria to face a new challenge: the chemo-mechanical stability of solid-solid interfaces. The quality and stability of this interface is key in enabling high performance solid-state batteries. Unlike conventional Li-ion batteries, new assembling processes are required to ensure the Li diffusion between the electrode and the solid electrolyte, using high pressure and/or temperature, which can lead to also reactivity between solid phases of different chemical nature. [1]-[2] Another completely new challenge is to deal with the mechanical constraints due to volume variation of electrode materials during charge and discharge, causing the disconnection between electrode and electrolyte materials, or formation of cracks, both effects will contribute to increase the cell resistivity and eventually to the cell failure. [3]-[4] This issue is generally mitigated by applying external pressure to the cell stack during cycling to actively maintain solid-solid interfaces. However, this solution strongly depends on the mechanical properties of the solid electrolyte material and introduces additional complexity that may hinder practical implementation and commercialization.

Alternatively, identifying electrode materials with little or no volume variation, also called “pseudo zero-strain” and “zero-strain” materials, can considerably decrease the pressure required during cycling, simplifying industrial developments and increasing the cycle life of solid-state batteries. A typical zero-strain material is the spinel lithium titanate ($Li_4Ti_5O_{12}$,, LTO), which does not change volume upon delithiation/lithiation and can sustain fast charging rates. [5] This material is widely used as negative electrode in high power Li-ion cells despite having a lower energy density compared to graphite-based electrode. Although no positive electrode has been reported with equivalent zero-strain properties, several studies focus in identifying such materials. [6]-[10] For example, Co-rich $LiNi_xMn_yCo_{1-x-y}O_2$ (NCM) compositions, such as NCM361 and NCM271, have been reported to exhibit smaller volume variations than their Ni-rich counterparts. [7] Lithiated spinel $LiCo_{1-x}Al_xO_2$ has also been associated with exceptionally low volume variation, among the lowest values reported for a zero-strain Li-ion positive electrode. [6]

In this context, understanding the volume variation of electrode materials upon cycling is particularly important for identifying promising positive electrode materials for ASSBs. To further guide the design of materials with controlled volume changes, it is essential to understand their structural evolution at the atomic scale and to establish how local structural changes contribute to the macroscopic volume variation observed during cycling. This requires complementary structural characterization techniques, such as X-ray diffraction (XRD) and X-ray absorption spectroscopy (XAS), which provide access to the average long-range crystal structure and the local atomic structure, respectively. Owing to its chemical sensitivity, XAS can also provide information on the redox mechanisms of transition metal (TM) oxide materials during cycling, thereby providing further insights into the underlying electrochemical mechanisms.

In particular, Co-substituted $LiCo_xMn_{2-x}O_4$ spinel material shows significantly reduced volume variations during cycling, together with an increased operating potential associated to redox activity of Co. [11]-[14] Although these materials cannot be strictly considered zero-strain compounds, they represent an interesting case study for investigating the relationship between local and average volume variations. Their average volume variation has been reported to vary with the Co/Mn ratio [11], while the charge compensation mechanism due to two redox centers (Co/Mn) can be probed by X-ray absorption spectroscopy (XAS). These features make the $LiCo_xMn_{2-x}O_4$ ($0 \leq x \leq 1$) spinel system a promising platform for investigating the effect of transition-metal substitution on the structural evolution of electrode materials.

In this work, we combine *operando* XRD and XAS to investigate the volume variation of spinel $LiCo_xMn_{2-x}O_4$ across the solid-solution series. Five compositions (x=0, 0.25, 0.5, 0.75, 1.0) were synthesized and electrochemically characterized, among which three representative compositions (x = 0, 0.5, and 1.0) were selected for the detailed investigation of their volume variation using *operando* XRD and XAS. To ensure reliable and comparable results from both *operando* techniques and to preserve the correlative nature of the study, we used the same electrochemical in situ cell and cycling protocol for both experiments. This enables us to access reliable information on the bulk structural evolution of the material, providing a basis for correlating structural changes with its electrochemical behavior. Finally, this correlative approach offers a framework for disentangling the structural response of distinct redox-active cations in solid-solution electrode materials, providing a methodological basis for future studies aiming to correlate composition, redox mechanism, and structural evolution, with relevance to advancing current Li-ion battery technology toward to reach solid-state architecture.

# 1. Results and Discussions

## 1.1. Structural characterization of $LiCo_xMn_{1-x}O_4$ spinel phases

The powder XRD patterns of the five synthesized $LiCo_xMn_{2-x}O_4$ compositions (x = 0, 0.25, 0.50, 0.75, and 1.0) are presented in **Figure 1a**. All compositions index consistently to the cubic spinel structure with space group $Fd\bar{3}m$, confirming the successful incorporation of Co into the spinel lattice across the full substitution range without inducing a phase transformation. The diffraction patterns exhibit the characteristic reflection set expected for this structure type, with the dominant 111 reflection.

A weak additional reflection at 2θ ≈ 18.6° emerges for the Co-rich compositions and is particularly evident for $LiCoMnO_4$ that may indicate the formation of $Li_2MnO_3$. The presence of such a phase has previously been reported in the synthesis of $LiCoMnO_4$ materials and has been attributed to a disproportionation process occurring following oxygen loss from the spinel structure. [12]-[15] Another reflection at 2θ ≈ 32° is also observed for the Co-rich compositions and becomes particularly evident for x = 0.75 and 1. This can be attributed to the partial occupation of tetrahedral 8*a* Li sites by cobalt [11],[16]. Rietveld refinement shows that this Co migration remains marginal for $LiCo_{0.75}Mn_{1.25}O_4$, but becomes much more significant for $LiCoMnO_4$ with 16 % of tetrahedral sites occupied by cobalt cations.

The progressive substitution of Mn by Co is accompanied by a systematic shift of the 111 reflection toward higher 2θ values, as shown in **Figure 1b**. This shift indicates a contraction of the spinel unit cell, consistent with the smaller ionic radii of $Co^{4+}/Co^{3+}$ compared with $Mn^{4+}/Mn^{3+}$ in octahedral coordination. Accordingly, Rietveld refinement of the XRD patterns reveals a systematic decrease in the lattice parameter *a* from $LiMn_2O_4$ to $LiCoMnO_4$ with increasing Co content (**Table 1**, **Figure S1**). The continuous evolution of the lattice parameter across the compositional series is consistent with the progressive incorporation of Co into the spinel framework, indicating the formation of a homogeneous solid solution over the investigated composition range.

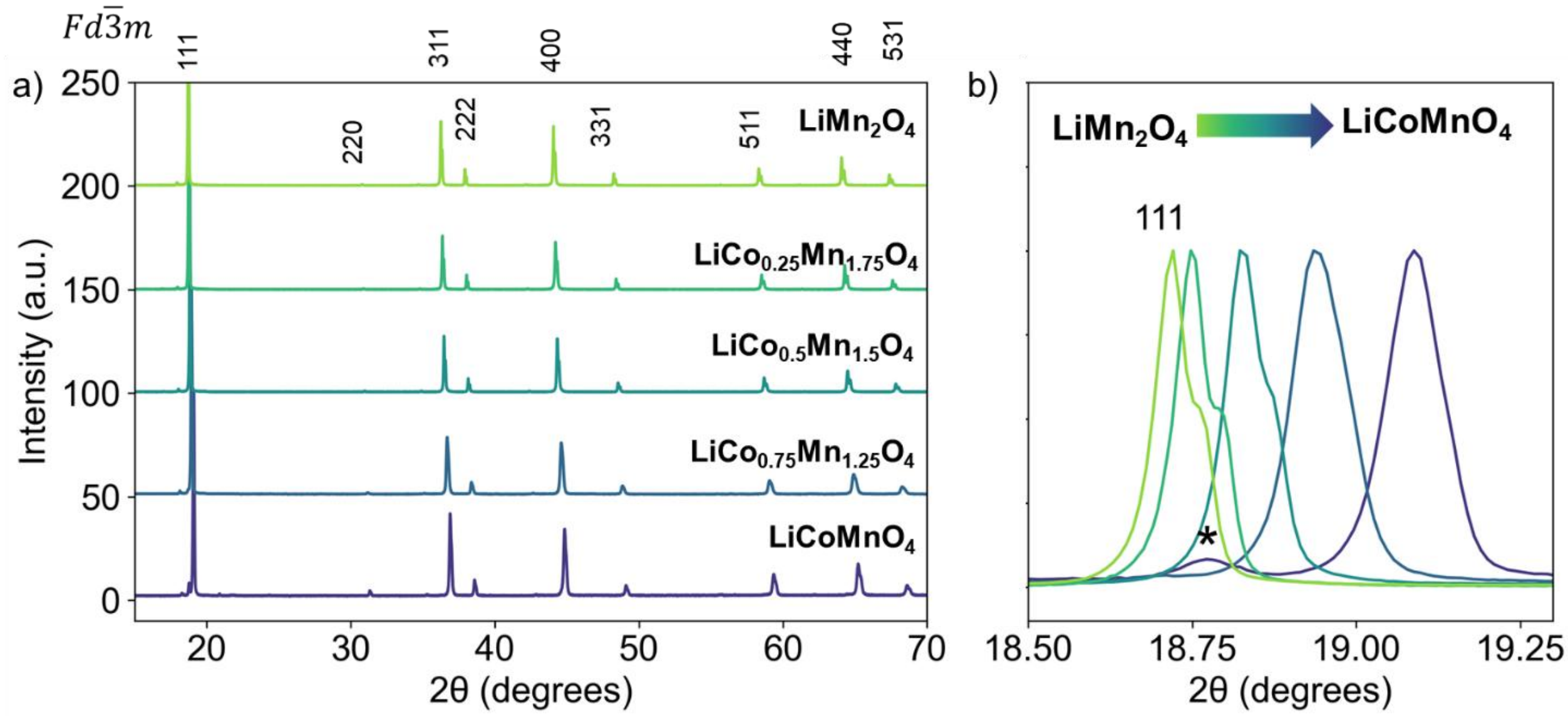

**Figure 1.** a) XRD patterns of $LiCo_xMn_{2-x}O_4$ sample (x=0, 0.25, 0.50, 0.75, 1.0). b) Zoom of the (111) peak exhibiting a shift to higher 2θ as the cobalt concentration increases. Small impurity of $Li_2MnO_3$ is present for higher Co content $LiCoMnO_4$ (denoted by *).

To complement the structural analysis, the morphology and elemental distribution of the synthesized materials were investigated by SEM coupled with EDX (**Figure S2**). The as-synthesized powders exhibit a homogeneous morphology, with an average particle size between 3 and 5 µm. Higher-magnification SEM images reveal small particles with polyhedral shapes and smooth surfaces. The small particle size and rounded edges are likely related to the low annealing temperature during synthesis. **Error! Reference source not found.** EDX elemental mapping confirms a homogeneous distribution of Mn and Co throughout the particles for most compositions. However, $LiCoMnO_4$ exhibits a few Mn-rich, Co-deficient regions, suggesting local compositional heterogeneity or the presence of a minor $Li_2MnO_3$ impurity. This may explain the slight deviation of $LiCoMnO_4$ from Vegard's law (**Figure S1 a**).

**Table 1.** Structural parameters for the synthesized $LiCo_xMn_{2-x}O_4$ samples with x = 0, 0.25, 0.5, 0.75, 1 determined by Rietveld refinement with space group $\boldsymbol{Fd\bar{3}m}$ for all the components.

| x | *a* (Å) | Atom | Wyckoff position | Occupancy | Atomic Coordinates | | |
|---|---|---|---|---|---|---|---|
| | | | | | x | y | z |
| **0** | 8.22434(17) | Li | 8*a* | 1 | 0.125 | 0.125 | 0.125 |
| | | Mn | 16*d* | 1 | 0.5 | 0.5 | 0.5 |
| | | O | 32*e* | 1 | 0.26482(11) | 0.26482(11) | 0.26482(11) |
| **0.25** | 8.19405(2) | Li | 8*a* | 1 | 0.125 | 0.125 | 0.125 |
| | | Co2 | 16*d* | 0.125 | 0.5 | 0.5 | 0.5 |
| | | Mn | 16*d* | 0.875 | 0.5 | 0.5 | 0.5 |
| | | O | 32*e* | 1 | 0.2647(1) | 0.2647(1) | 0.2647(1) |
| **0.5** | 8.14644(2) | Li | 8*a* | 1 | 0.125 | 0.125 | 0.125 |
| | | Co2 | 16*d* | 0.25 | 0.5 | 0.5 | 0.5 |
| | | Mn | 16*d* | 0.75 | 0.5 | 0.5 | 0.5 |
| | | O | 32*e* | 1 | 0.26537(8) | 0.26537(8) | 0.26537(8) |
| **0.75** | 8.09893(3) | Li | 8*a* | 0.983 | 0.125 | 0.125 | 0.125 |
| | | Co1 | 8*a* | 0.017 | 0.125 | 0.125 | 0.125 |
| | | Co2 | 16*d* | 0.488 | 0.5 | 0.5 | 0.5 |
| | | Mn | 16*d* | 0.512 | 0.5 | 0.5 | 0.5 |
| | | O | 32*e* | 1 | 0.26286(8) | 0.26286(8) | 0.26286(8) |
| **1** | 8.07514(4) | Li | 8*a* | 0.841 | 0.125 | 0.125 | 0.125 |
| | | Co1 | 8*a* | 0.159 | 0.125 | 0.125 | 0.125 |
| | | Co2 | 16*d* | 0.512 | 0.5 | 0.5 | 0.5 |
| | | Mn | 16*d* | 0.488 | 0.5 | 0.5 | 0.5 |
| | | O | 32*e* | 1 | 0.26470(9) | 0.26470(9) | 0.26470(9) |

### 1.2. Electrochemical properties

The galvanostatic charge-discharge profiles of $LiCo_xMn_{2-x}O_4$ as a function of Co substitution are presented in **Figure 2**. A clear and systematic evolution of the electrochemical response is observed with increasing Co content. In $LiMn_2O_4$ (x = 0), the electrochemical activity is dominated by the two characteristic voltage plateaus of the $Mn^{4+}/Mn^{3+}$ redox couple in the 3.9-4.2 V region.

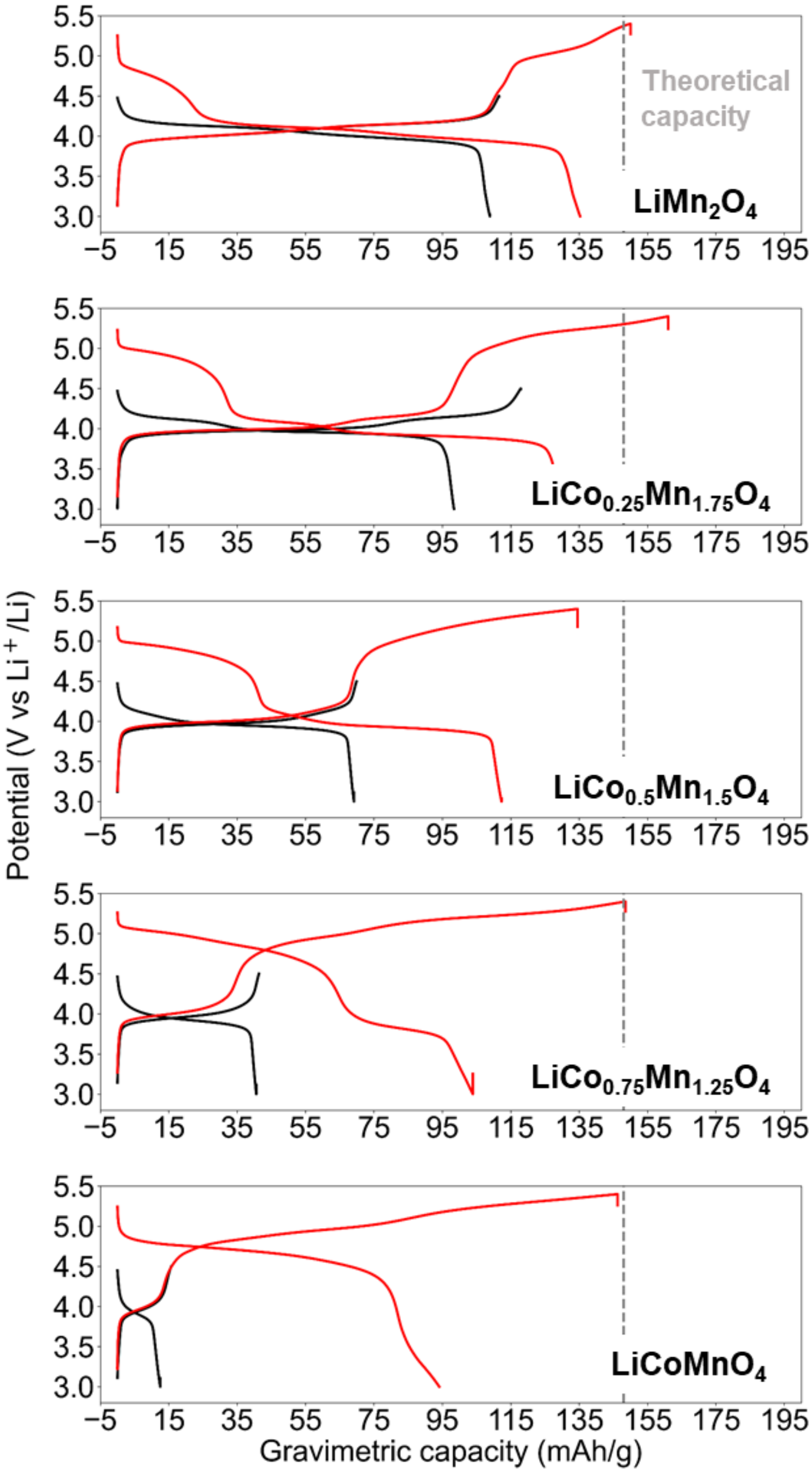

**Figure 2.** Electrochemical profiles of $LiCo_xMn_{2-x}O_4$ half cells at different Co substitution level. Dashed lines indicate the theoretical capacity expected based on one lithium ion exchanged per formula unit. Black curves represent the first cycle limited at 4.5 V during charge, while red curves represent the second cycle reaching 5.4 V during charge.

As Co content increases, the lower voltage plateaus are progressively suppressed, while a new high-voltage plateau emerges and grows, consistent with the activation of the $Co^{4+}/Co^{3+}$ redox couple above ~4.7 V. In $LiCoMnO_4$ (x = 1.0), almost 90% of the capacity arises from this high-voltage process, confirming the dominant role of Co redox activity in the fully substituted composition. These electrochemical profiles confirm the activation of the relevant redox processes and define the accessible voltage windows for each composition, which guided the potential limits selected for the subsequent *operando* XRD and XAS measurements.

### 1.3. Charge compensation mechanism by *operando* XANES

XAS on *ex situ* electrode and in *operando* conditions was applied to provide complementary insight into the structural and electronic evolution of cathode materials, at selected states of charge as well during electrochemical cycling. Although *ex situ* XAS studies have previously been conducted, *ex situ* measurements can be affected by structural relaxation or self-discharge during cell disassembly and sample handling. [18]-[20]

Therefore, Mn and Co K-edge XAS spectra were collected from *ex situ* electrodes prepared at selected states of charge (pristine, charged at 4.5 V, charged at 5.4 V, discharged at 4.5 V, and discharged at 3 V), and evaluated alongside spectra obtained under *operando* conditions in half-cell configuration. Comparison of the X-ray near edge structure (XANES) spectra at high voltage (**Figure S3**) reveals clear differences between the *ex situ* and corresponding *operando* measurements, particularly in the edge position and pre-edge feature shapes. Based on edge shift analysis, the *ex situ* samples for all three compositions ($LiMn_2O_4$, $LiCo_{0.5}Mn_{1.5}O_4$, and $LiCoMnO_4$) exhibited noticeably lower oxidation states than their *operando* counterparts for both Co and Mn K-edges. This discrepancy likely arises from structural and electronic relaxation of the positive electrode upon removal from the electrochemical cell potentially accompanied by partial self-discharge during cell disassembly and sample handling. Corroborating this observation, a previous *ex situ* XAS study of $LiCo_{0.5}Mn_{1.5}oO_4$ [18] reported a relatively small energy shift at the Co K-edge (~ 0.3 eV) compared to the 1.0 eV shift obtained in our *operando* study. This difference suggests that relaxation or self-discharge may partially obscure the oxidation-state changes occurring during lithiation and delithiation cycles for *ex situ* measurements. Similarly, another *ex situ* study on $LiCoMnO_4$ [20] reported no shift in the Mn K-edge, whereas our *operando* measurements revealed a shift of ~ 0.4 eV. These findings underscore the importance of *operando* measurements for obtaining an accurate and representative picture of the charge compensation mechanisms of the electrochemical processes, especially at high voltage [21]-[22].

*Operando* XAS measurements were performed on three selected $LiCo_xMn_{2-x}O_4$ compositions (x=0, 0.5, 1) using the same in situ cell for XRD (see **section 4.4**). Details about the electrochemical protocol are reported in **Table S1**. For $LiMn_2O_4$ the redox activity is only ascribed to activation of the $Mn^{4+}/Mn^{3+}$ couple below approximately 4.5 V, higher potential resulting in serious electrolyte degradation and no additional reversible capacity. In contrast, $LiCo_{0.5}Mn_{1.5}O_4$ and $LiCoMnO_4$ were cycled to higher voltages (5.3–5.4 V) to access the $Co^{4+}/Co^{3+}$ redox process. For $LiCo_{0.5}Mn_{1.5}O_4$, a C/2 (339 μA $cm^{-2}$) rate was required to reach the high-voltage plateau at 5.4 V, since at high potential the delithiation reaction competes with electrolyte oxidation and its associated re-intercalation, at the slower C/5 (219 μA $cm^{-2}$) rate this competition limits the cell to lower voltages (**Figure S4, S5**). Consequently, this study focuses on the $LiCo_{0.5}Mn_{1.5}O_4$ cell cycled at C/2, while the other two compositions were cycled at C/5. For LiCoMnO4, the cell was charged to a target of 5.4 V but reached only 5.3 V, cycling was stopped at this point to limit further electrolyte degradation at high potential (**Figure S6**).

The electrochemical profiles along with the corresponding *operando* XANES spectra for the three positive electrodes are shown in **Figure 3**. As the absorption-edge energy of a transition metal shifts toward higher energy with increasing its oxidation state, we have monitored the edge position to track evolution of Mn and Co oxidation state to evaluate their respective contributions to charge compensation during cycling.

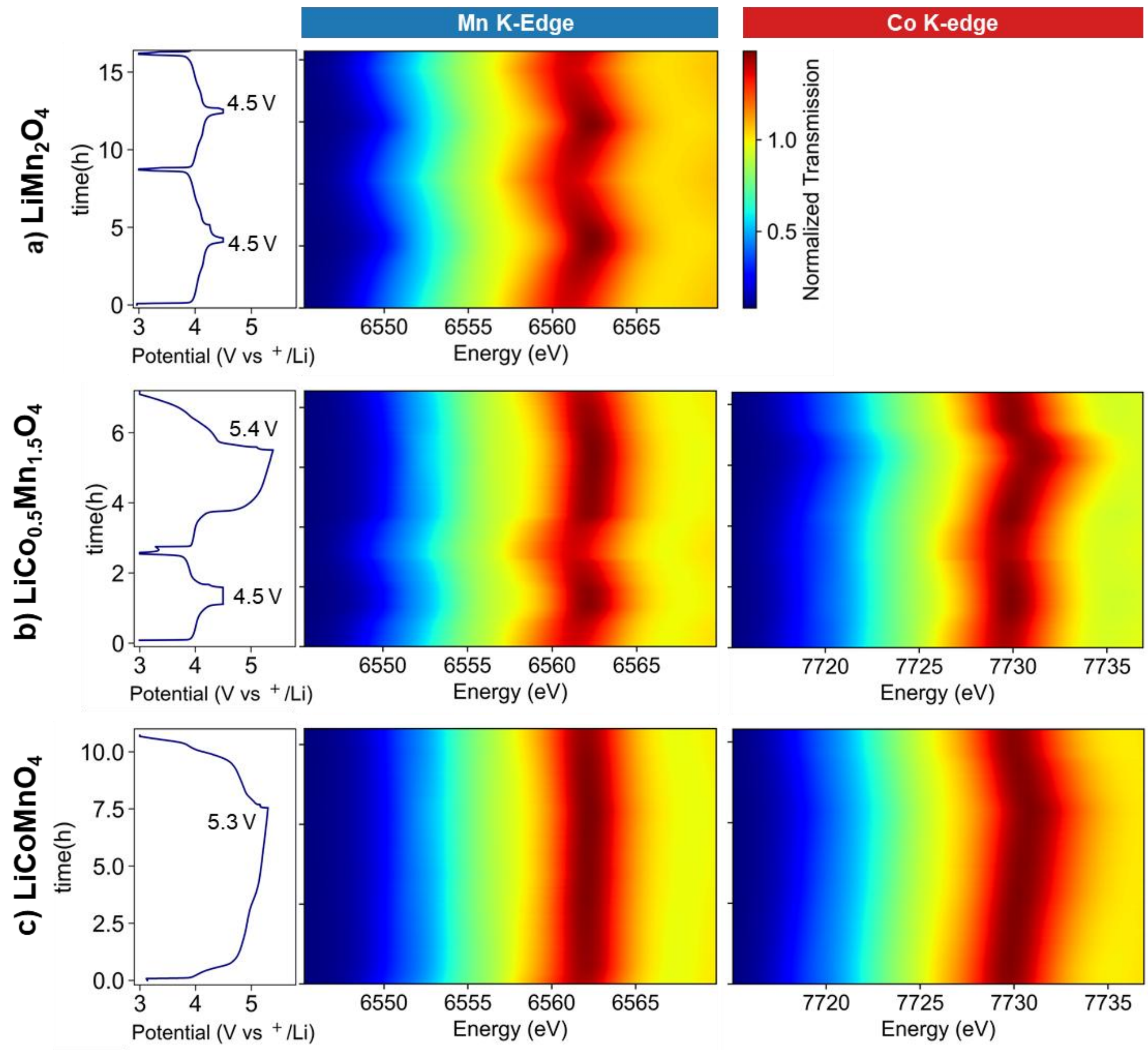

**Figure 3.** Electrochemical profile (left column) along with the corresponding normalized *operando* XANES spectra at the Mn K-edge (middle column) and Co K-edge (right column) for the three cathode materials. The $LiMn_2O_4$ cell (a) was charged at C/5 (176 µA $cm^{-2}$), the $LiCo_{0.5}Mn_{1.5}O_4$ cell (b) was charged at C/2 (339 µA $cm^{-2}$), and the $LiCoMnO_4$ cell (c) was charged at C/5 (133 µA $cm^{-2}$).

The edge energies used to assess the redox evolution of Mn and Co were determined from the normalized absorption spectra at J = 1.0 for the Mn K-edge and J = 0.8 for the Co K-edge. The XANES spectra of reference compounds are shown in the Supplementary Information (**Figure S8, S9**). **Figure 4** shows the evolution of the Mn and Co edge position during charge and discharge, together with the corresponding galvanostatic charge-discharge (GCPL) profiles. The edge shift toward higher energy reflects oxidation of the transition metal, while a shift toward lower energy indicates its reduction.

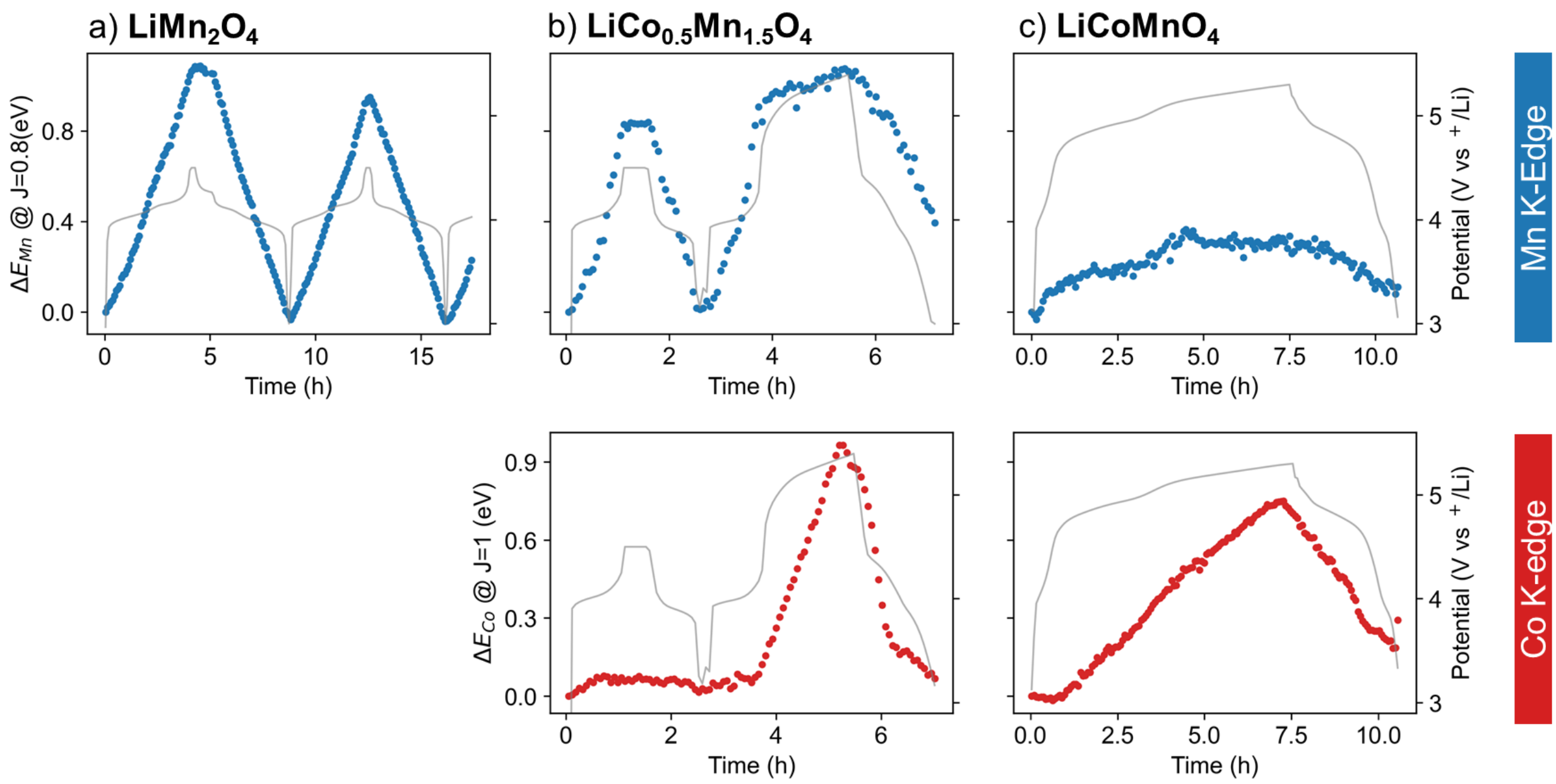

**Figure 4.** Energy shifts calculated for Mn K-edge (top panel, blues dots) and Co (bottom panel, red points) *operando* XANES spectra during cycling (grey curve) for a) $LiMn_2O_4$, b) $LiCo_{0.5}Mn_{1.5}O_4$, and c) $LiCoMnO_4$.

For $LiMn_2O_4$, the average Mn oxidation state in the pristine material was determined from the Mn K-edge position via a linear calibration curve constructed from $Mn^{3+}$ and $Mn^{4+}$ reference compounds (**Figure S8**) and it is approximately +3.5, consistent with the expected composition. Upon charging up to 4.5 V vs $Li^+$/Li, Mn is progressively oxidized, reaching an average oxidation state close to +4 at the end of charge (**Figure 4 a**). During the subsequent discharge, the Mn oxidation state decreases and returns to a value close to that of the pristine state, indicating a largely reversible redox process. A similar evolution is observed during the second electrochemical cycle. However, a slight degree of irreversibility is evidenced by the lower oxidation state reached at the end of the second charge compared to that of the first charge.

The Co substitution into the spinel structure results in the activation of the $Co^{4+}/Co^{3+}$ redox couple at higher potentials, with the extent and onset potential of this contribution depending on the degree of Co substitution. For the partially substituted composition $LiCo_{0.5}Mn_{1.5}O_4$, the cell was charged to 4.5 V during the first cycle and then up to 5.4 V during the second cycle, at a rate process remains overall highly reversible over of C/2 (339 μA $cm^{-2}$). The evolution of the Mn and Co K-edge positions reveals distinct redox behaviors at different voltages. Below 4.5 V, only Mn is electrochemically active (**Figure 4 b, top**), with no noticeable change in the Co oxidation state (**Figure 4 b, bottom**). During the second cycle, charged up to 5.4 V, the Co redox process becomes active at approximately 4.7 V (**Figure 4 b, bottom**), corresponding to the oxidation of $Co^{3+}$ to $Co^{4+}$. Thus, $LiCo_{0.5}Mn_{1.5}O_4$, Mn oxidation dominates the early stage of charge, while Co oxidation only sets in at higher potentials, above ~4.7 V.

For the fully Co-substituted composition, $LiCoMnO_4$, charge compensation is expected to be carried almost entirely by the Co redox couple, with Mn nominally close to the +4 state. However, a short low-voltage plateau is still observed during cycling, indicating that a small fraction of Mn is not purely tetravalent in the pristine material and remains electrochemically active. Consistent with this, a slight energy shift is still observed at the Mn K-edge. This deviation suggests that the Mn electronic configuration is influenced by local charge-compensation mechanisms rather than being electrochemically inert as formally expected. In $LiCoMnO_4$-based spinels, oxygen non-stoichiometry and structural defects, such as oxygen vacancies and cation disorder, have been reported to induce partial reduction of $Mn^{4+}$ toward $Mn^{3+}$, modifying the local electronic structure and electrochemical activity of the material [12]-[14]. This pre-existing $Mn^{3+}$ population, in turn, provides a small additional pool of Mn available for oxidation during charge, consistent with the slight edge shift observed. However, the precise contribution of these defect-driven effects versus an intrinsic electronic redistribution mechanism remains unclear and requires further investigation. Overall, the correspondence between the edge-shift magnitude and the capacity delivered at each voltage window indicates that the capacity in this system arises solely from cationic redox activity of the transition metals, with no evidence of anionic (oxygen) redox contribution.

The evolution of the XANES edge positions therefore reveals a voltage-dependent charge-compensation mechanism in $LiCo_xMn_{2-x}O_4$, in which the two transition-metal redox centers become active at distinct potentials. Mn oxidation dominates at lower voltages, below approximately 4.7 V, whereas Co oxidation becomes significant at higher voltages above this potential. This is consistent with the evolution of the electrochemical profiles observed in the coin cell, validating our approach (**Figure 2**). As the Co content increases, the lower-voltage $Mn^{4+}/Mn^{3+}$ plateau progressively decreases and is replaced by the higher-voltage $Co^{4+}/Co^{3+}$ plateau, reflecting the increasing contribution of Co to charge compensation. This tandem activation of Mn and Co redox centers is directly resolved by *operando* XAS measurements, whereas the corresponding redox processes are substantially less apparent in the *ex situ* measurements presented in **Figure S3** and in previous studies. [18],[20] Moreover, *operando* XAS measurements provide access not only to the electronic evolution through XANES analysis but also to the accompanying local structural changes through EXAFS modeling, identifying the modification of TM - oxygen and TM - TM bond lengths and their correlation with the changes in the unit-cell volume during cycling.

### 1.4. Structural evolution from correlative *operando* XAS and XRD

Understanding the structural evolution of electrode materials during electrochemical cycling requires complementary insights into both long-range crystallographic changes and local atomic rearrangements. *Operando* XRD and XAS provide powerful complementary approaches to monitor these structural responses in real time under working conditions. While XRD tracks the evolution of the average crystal structure through changes in lattice parameters and unit cell volume, XAS provides element-specific information on the local coordination environment, including variations in bond distances and coordination that may remain hidden in diffraction measurements, particularly for TM elements with similar X-ray scattering factors. By correlating these two structural probes, it becomes possible to establish a comprehensive picture of the lattice response during lithium extraction and reinsertion, linking macroscopic structural changes with the underlying local rearrangements occurring around the TM sites.

For XRD, the unit cell volume evolution was obtained from the refined lattice parameter *a*. The unit cell volume can also be extracted from EXAFS through the TM-TM bond length, $d_{TM-TM}$, which relates to the lattice parameter *a* via the crystallographic relationship illustrated in **Figure 5**. In the spinel structure, Li occupies the tetrahedral 8*a* sites (green spheres), while Mn and substituted Co atoms occupy the octahedral 16*d* sites (purple spheres), and oxygen resides on the 32*e* sites (red spheres). Since the face diagonal of a cube with edge ***a*** has length $a\sqrt{2}$, the nearest TM-TM distance $d_{TM-TM}$ is

**Equation 1** $$d_{TM-TM} = \frac{a\sqrt{2}}{4}$$

Rearranging this to solve for *a*:

**Equation 2** $$a = \frac{4d_{TM-TM}}{\sqrt{2}} = 2\sqrt{2} \cdot d_{TM-TM}$$

giving the geometric relationship between $d_{TM-TM}$ (measured by EXAFS) and ***a*** (measured by XRD) used to compute the XAS-derived cell volume.

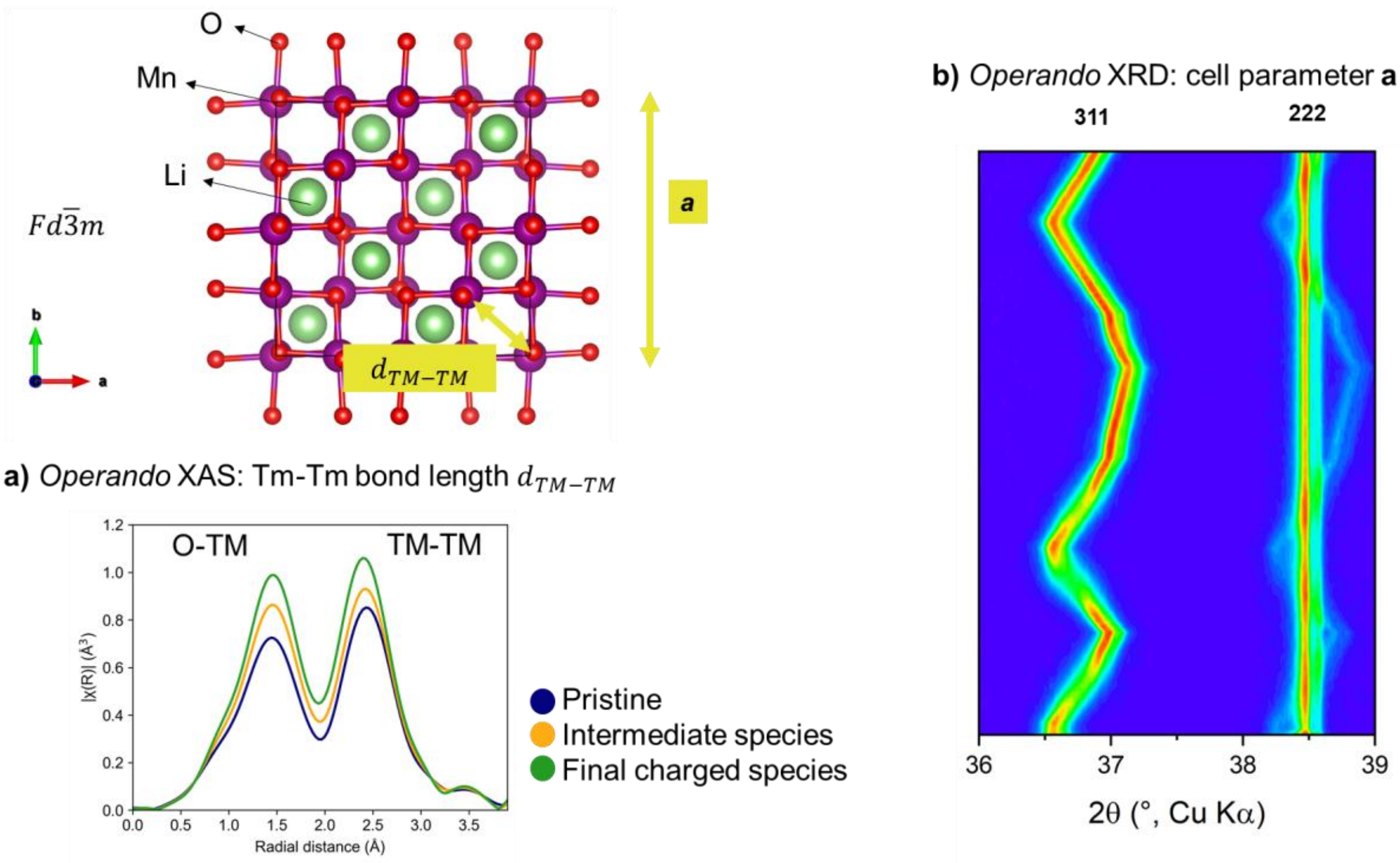

**Figure 5.** Geometrical relationship the cubic lattice parameter *a* and the transition metal-transition metal (TM-TM) bond length $d_{TM-TM}$ in $LiMn_2O_4$ materials during cycling. (a) Fourier transform of the EXAFS data from operando XAS, showing the TM-TM bond length. (b) *Operando* XRD data, from which the lattice parameter *a* is obtained, represented by the (311) and (222) reflections. The intense, unchanging pak near 38.5° originates from the Al current collector of the operando cell.

### 1.4.1. Long range structure by *operando* XRD

All three cells were cycled for two charge-discharge cycles at a rate of C/5 using the voltage windows shown in **Figure 6 b** (gray curve, right y-axis). Although $LiCo_{0.5}Mn_{1.5}O_4$ requires a charging rate of C/2 to reach the upper cut-off voltage of 5.4 V in this *operando* cell configuration, as discussed in **Section 2.3** and **Figure S4**, a C/5 rate was employed here to provide sufficient acquisition time for each diffraction pattern while maintaining consistent experimental conditions across all compositions. Consequently, all cells were limited at an upper cut-ff voltage of 5.3 V.

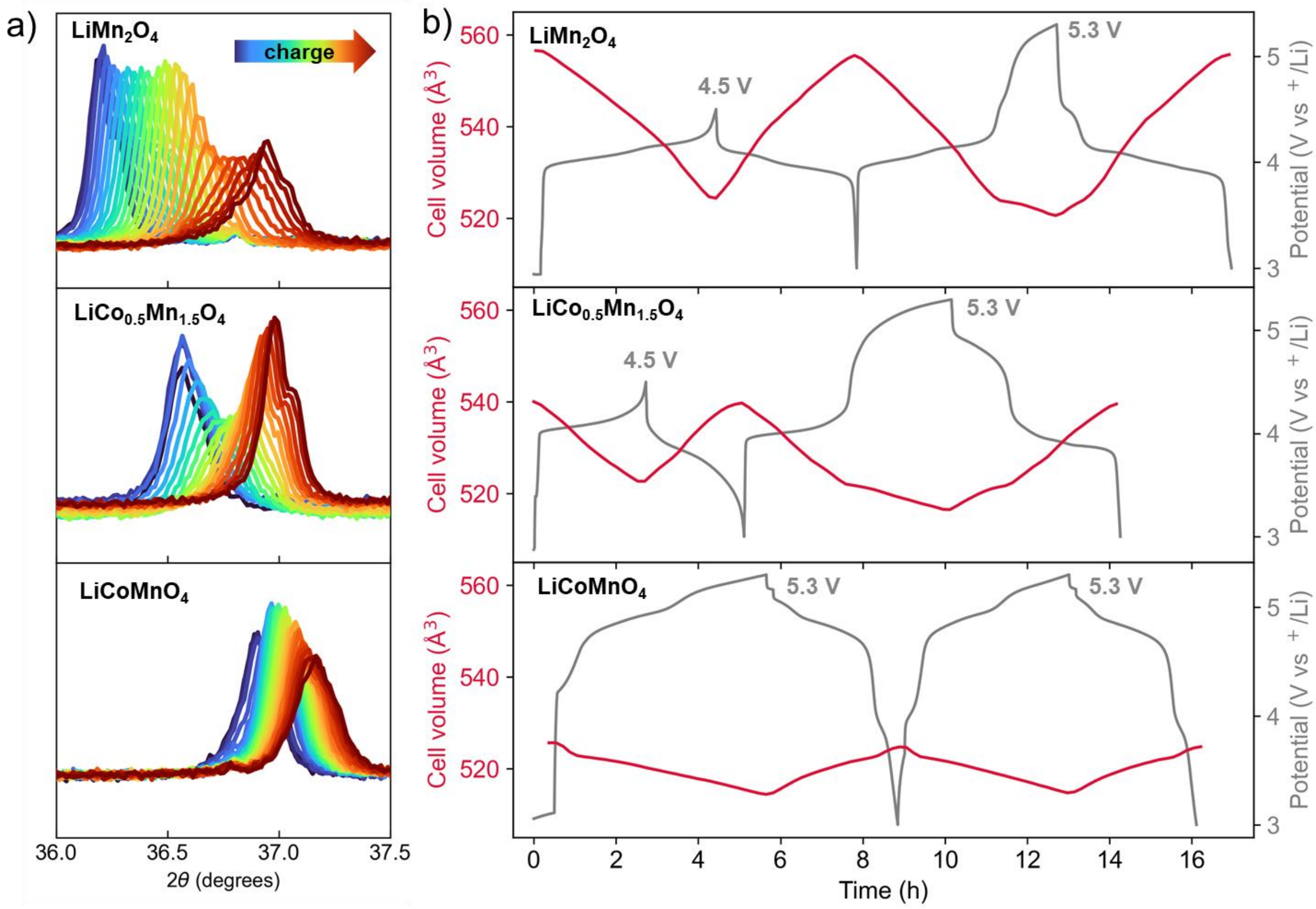

**Figure 6.** a) Evolution of the 311 peak of the spinel structure for $LiMn_2O_4$ (top), $LiCo_{0.5}Mn_{1.5}O_4$ (middle), and $LiCoMnO_4$ (bottom) during first charge, obtained from *operando* XRD measurements. b) Electrochemical profile of the materials during cycling (black curve) and the corresponding volume changes (red curve).

**Figure 6 a** focuses on the evolution 311 peak during the first charge for all materials, showing a change of *a* lattice parameter due to lithium extraction. Cell volume values from Rietveld fitting of the XRD pattern during charge and discharge are shown in **Figure 6 b** (red curve, right y-axis). Indeed, as the voltage increases during charge, the cell volume decreases as a result of lattice contraction associated with TM oxidation upon delithiation, as reflected by the shift of the diffraction peaks toward higher 2θ values. Conversely, the cell volume expands during discharge as lithium is reinserted into the structure. For the lower-voltage redox processes, such as the first cycle of $LiMn_2O_4$ and $LiCo_{0.5}Mn_{1.5}O_4$, the evolution of the cell volume is nearly linear with voltage. In contrast, at higher voltages we observe a break in the cell-volume variation, indicating the onset of the $Co^{4+}/Co^{3+}$ redox reaction. This nonlinear response reflects the distinct structural evolution associated with the high-voltage cobalt redox process compared with the low-voltage $Mn^{4+}/Mn^{3+}$ redox process as discussed in **Section 2.3**.

### 1.4.2. Local structure by *operando* EXAFS

The ability of XAS to accurately differentiate the Co-O and Mn-O as well as the Co(Mn)-Mn(Co) interatomic distances makes it a powerful technique for probing the local atomic environment during electrochemical cycling around the TM sites.

Chemometric analysis tools, specifically Principal Component Analysis (PCA) and Multivariate Curve Resolution with Alternating Least Squares (MCR-ALS), were employed to extract chemically meaningful information from the large XAS dataset. PCA performed on the full XAS spectral dataset indicates that three independent components are sufficient to describe the electrochemical processes occurring across all compositions. MCR-ALS was subsequently used to decompose the XAS datasets into the three orthogonal component spectra and their corresponding concentration profiles throughout electrochemical cycling. These three components are: (i) the species at open circuit voltage (OCV), referred to hereafter as the "*Pristine*" species; (ii) a transient species emerging during the charge and never reaching to 100% concentration, referred to as the "*Intermediate*" species; and (iii) a species that appears exclusively toward the end of charge, referred to as the "*Final Charged*" species. While this finding is common to all three materials upon delithiation, the corresponding reconstructed XAS spectra differ depending with the Co-content.

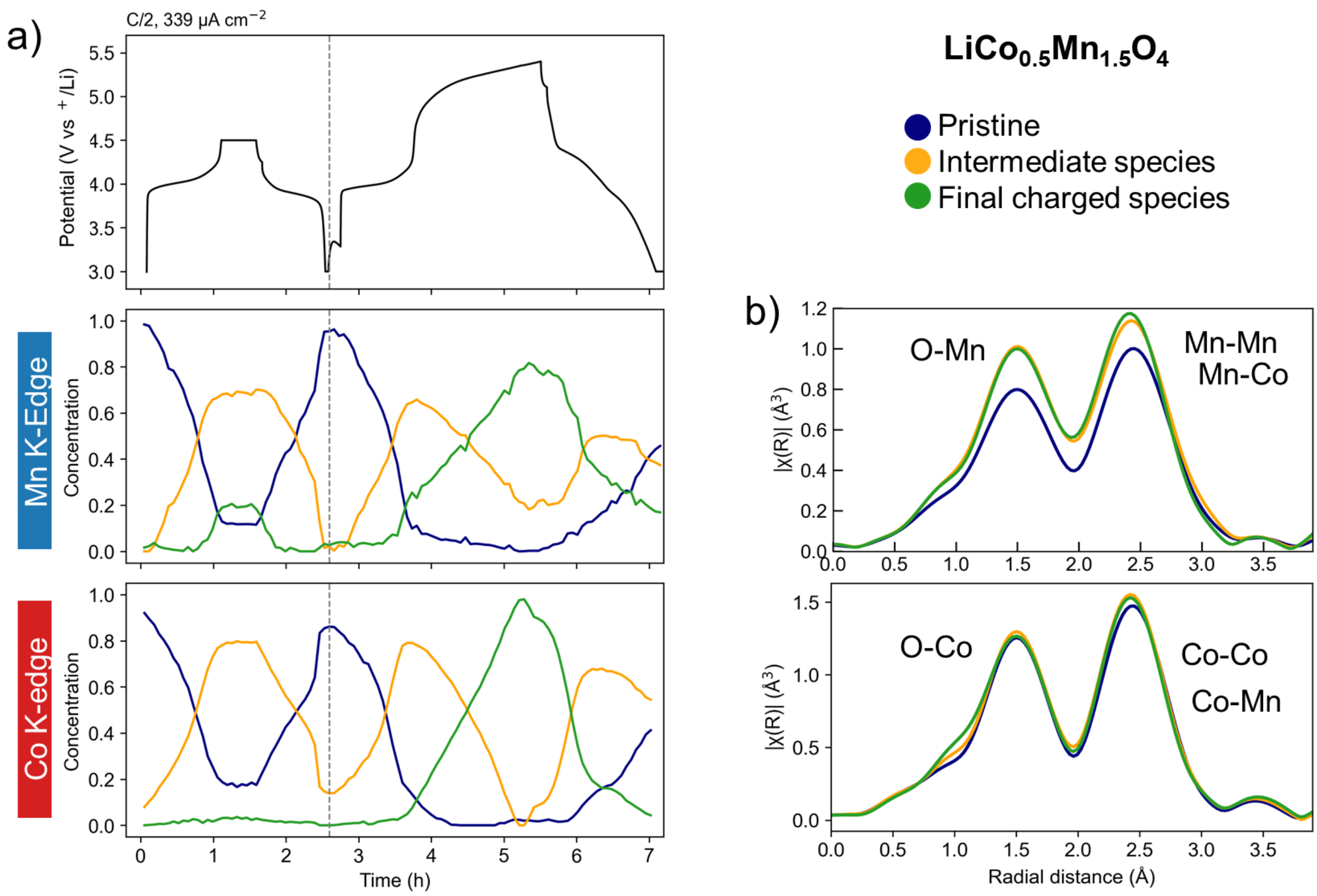

**Figure 7.** MCR-ALS analysis on $LiCo_{0.5}Mn_{1.5}O_4$ dataset results in three reconstructed principal components: *Pristine* (blue), *Intermediate species* (yellow), and *Final Charged species* (green). a) Concentration profile of these three components during cycling process (top panel, black line) at Mn K-edge (middle panel) and Co K-edge (bottom panel). b) Fourier transforms (FTs) of the EXAFS oscillations extracted at Mn K-edge (middle panel) and Co K-edge (bottom panel). The FTs are not corrected by the phase shifts.

The results obtained on $LiCo_{0.5}Mn_{1.5}O_4$ as a representative example (**Figure 7**). A detailed description of the results obtained by MCR-ALS analysis for all compositions is provided in the **Supporting Information** (**Section S2**). During the first stage of charge, the concentration of the *Pristine* (blue curve) progressively decreases as the *Intermediate* components (yellow curve) becomes dominant. Upon further delithiation toward the upper cut-off voltage of 5.4 V, the *Intermediate* species is gradually consumed and replaced by the *Final Charged* species (green curve), which becomes the dominant component at the end of charge.

To obtain further insight into the local structural evolution associated with the reconstructed spectral components, the EXAFS oscillations at both the Mn and Co K-edges were modeled. The Fourier transforms exhibit two dominant contributions: the first peak, located at lower radial distances, corresponds to the transition metal-oxygen (TM-O) coordination shell, whereas the second peak, at higher radial distances, arises from transition metal-transition metal (TM-TM) scattering paths.

The EXAFS spectra of the pristine materials at both the Mn and Co K-edges were refined using the spinel structure $Fd\bar{3}m$ $LiMn_2O_4$ as the initial structural model. For compositions containing Co, the transition metal site in the structural model was populated with a mixture of Mn and Co atoms according to the nominal Co:Mn occupancy of each composition, to accurately account for both absorbing and backscattering contributions at the two edges. Co was placed on the octahedral (16d) site in all compositions. To ensure consistency between both absorption edges, the transition metal - transition metal (TM-TM) scattering paths were refined using shared structural parameters, while the remaining coordination shells were allowed to vary independently. The complete fitting procedure is described in **Experimental Section 4.6**, and the refined structural parameters for $LiMn_2O_4$, $LiCo_{0.5}Mn_{1.5}O_4$, and $LiCoMnO_4$ are summarized in **Table S2, S3, S4**, respectively.

Minimal changes are observed in the transition metal-oxygen (TM-O) bond distances across all compositions at both the Mn and Co K-edges, signaling the robustness and stability of the metal-oxygen framework under varying states of charge (**Figure 7 a**). In the second coordination shell, corresponding to transition metal-transition metal (TM-TM) interactions, a slight contraction is observed moving from the *Pristine* species through the *Intermediate* species and toward the *Final Charged* species. The shrinkage of the TM-TM bond distance is consistently observed across all compositions and at both K-edges, suggesting a coherent structural response to the charge compensation processes occurring throughout the electrochemical cycling, albeit small.

Additionally, fitting results show a shorter Co-TM bond distance compared to Mn-TM that has been reported in previous crystallographic studies. [18],[24]-[25] This might come from the local contraction of the Co-centered octahedron following substitution of the larger Jahn-Teller-active $Mn^{3+}$ by the smaller, low-spin $Co^{3+}$ ion. [25] The reduced ionic radius and absence of Jahn-Teller distortion allow the surrounding oxygen framework to relax inward, which in turn decreases the distances to neighboring transition-metal cations. [18] These local structural relaxations are averaged out in XRD but are directly captured by the element-specific local probe provided by EXAFS.

To understand the effect of Co substitution on the structural evolution during electrochemical cycling, the unit cell volume of each composition was determined under *operando* conditions. For $LiMn_2O_4$, since the transition metal sublattice consists exclusively of Mn atoms, the Mn-Mn bond distance $d_{Mn-Mn}$ obtained from EXAFS fitting can be directly related to the lattice parameter *a* (**Equation 2**) through the crystallographic relationship between the nearest-neighbor TM-TM distance and the cubic spinel cell volume $V_{LiMn_2O_4}$ :

**Equation 3**

$$a = 2\sqrt{2} \cdot d_{Mn-Mn}$$

$$V_{LiMn_2O_4} = a^3 = \left(2\sqrt{2} \cdot d_{Mn-Mn}\right)^3$$

For $LiCo_{0.5}Mn_{1.5}O_4$, where Mn and Co randomly co-occupy the octahedral 16d sites with fractional occupancies determined by the nominal composition (75% Mn and 25% Co), the nearest-neighbour TM-TM coordination shell probed by EXAFS at either the Mn or Co K-edge does not correspond to a single distance. Instead, the measured EXAFS signal represents a statistical contribution from different TM-TM scattering pairs (Mn-Mn with bond length $d_{Mn-Mn}$, Mn-Co with bond length $d_{Mn-Co}$, and Co-Co with bond length $d_{Co-Co}$). Therefore, a composition-weighted average TM-TM distance $d_{Mn-TM}$ and $d_{Co-TM}$ was defined for each absorbing element:

**Equation 4**

$$d_{Mn-TM} = x_{Mn} d_{Mn-Mn} + x_{Co} d_{Mn-Co}$$

$$d_{Co-TM} = x_{Mn} d_{Co-Mn} + x_{Co} d_{Co-Co}$$

where $x_{Mn}$ and $x_{Co}$ are the fractional occupancies of Mn and Co on the 16d site for each composition ($x_{Mn}$ = 0.75, $x_{Co}$ = 0.25 for $LiCo_{0.5}Mn_{1.5}O_4$, $x_{Mn}$ = $x_{Co}$ = 0.5 for $LiCoMnO_4$). With the relationship with cubic lattice parameter *a* shown in **Equation 2**, the unit cell a and cell volume $V_{LiCo_xMn_{2-x}O_4}$ can be calculated as

**Equation 5**

$$a_{Mn-TM} = 2\sqrt{2} \cdot d_{Mn-TM}$$

$$a_{Co-TM} = 2\sqrt{2} \cdot d_{Co-TM}$$

$$V_{LiCo_xMn_{2-x}O_4} = (x_{Mn} a_{Mn-TM} + x_{Co} a_{Co-TM})^3$$

Bond length values and its fitting parameters are detailed in **Table S2-S4** and **Section 4.5**, respectively.

The calculated unit cell volume from EXAFS were then combined with the concentration profiles obtained from MCR-ALS to reconstruct the local structural evolution during electrochemical cycling. At each time point (corresponding to one acquired spectrum), the average cell volume was calculated by weighting the contribution of each reconstructed principal component according to its corresponding concentration profile:

**Equation 6**

$$V_{XAS}(t) = \sum_{i=1}^{3} C_i(t)V_i$$

where $V_{XAS}(t)$ is the reconstructed XAS-derived cubic cell volume at time $t$ (i.e., at the corresponding spectrum number), $C_i(t)$ is the normalized concentration contribution of component obtained from MCR-ALS at time $t$, and $V_i$ is the cubic cell volume associated with each reconstructed component, obtained from the EXAFS-derived TM-TM distance and the crystallographic relationships described above (**Equation 3-6**).

This approach enables the evolution of the EXAFS-derived local structural volume to be compared directly with the unit cell volume variation obtained from *operando* XRD that will be discussed further in the next section.

### 1.4.3. Volume variations on local and long range

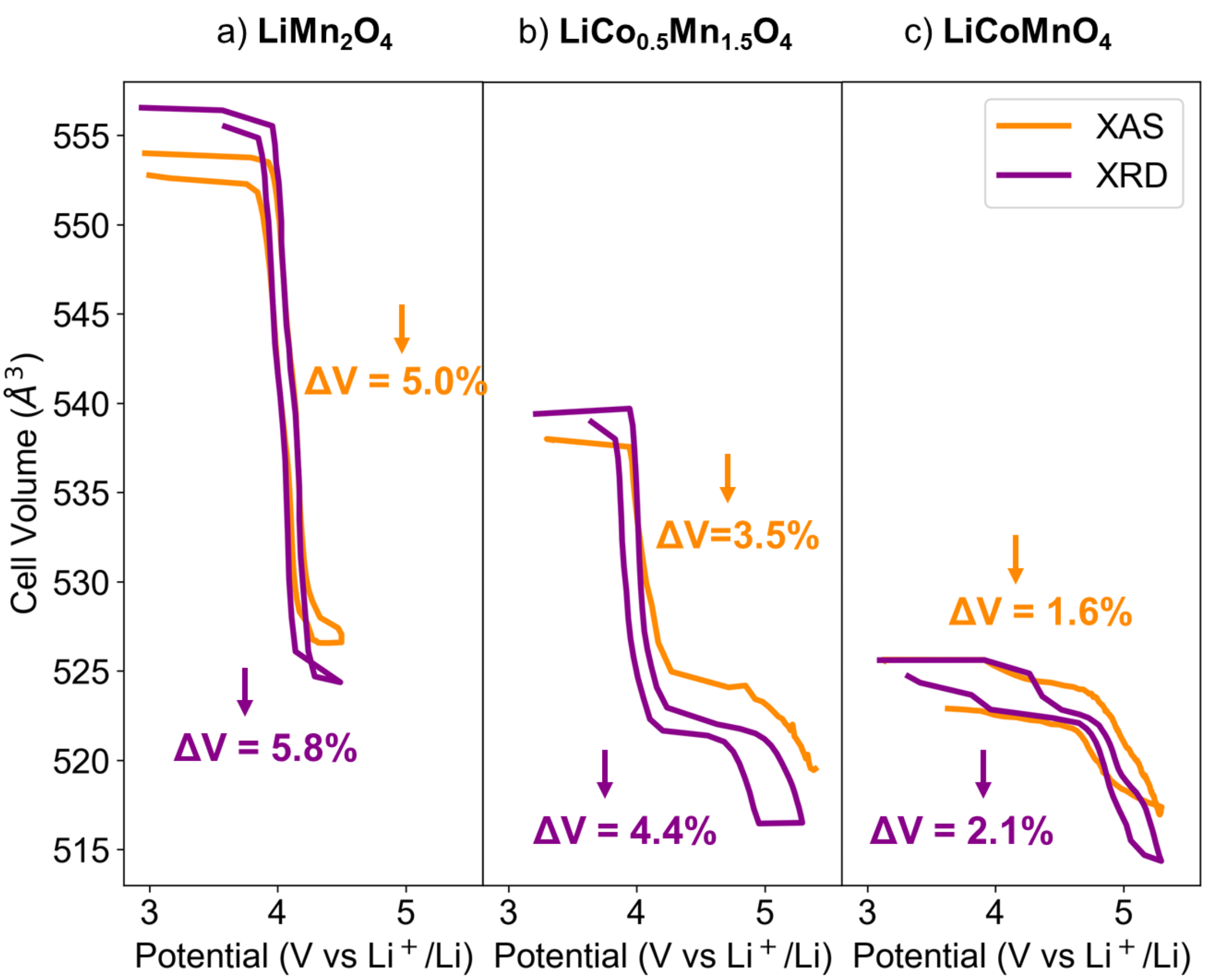

**Figure 8.** Comparison of volume variation obtained from *operando* XRD and XAS for the three compositions of the spinel $LiCo_xMn_{2-x}O_4$.

The unit cell volume evolution calculated from both techniques was compared as a function of potential (**Figure 8**) to represent the structural changes accompanying charge/discharge processes for each composition cycled at C/5. An exception applies to $LiCo_{0.5}Mn_{1.5}O_4$, for which the *operando* XAS and XRD measurements were conducted at different rates, C/2 and C/5, respectively. As established in **Section 2.3** and **Fig. S5-6**, a minimum rate of C/2 is required to drive this composition to the high-voltage plateau at 5.4 V under the *operando* cell configuration employed. However, the slower C/5 rate was necessary for the XRD measurements due to acquisition time constraints. Consequently, the discharge profile of the XAS cell for this composition is subject to significantly different electrochemical polarization relative to its XRD counterpart and has therefore been omitted from the comparison plot.

A good agreement between the unit cell volume evolution obtained from XRD and XAS is observed across the three compositions, as shown in **Figure 8**. For $LiMn_2O_4$, a volume contraction of approximately 5.8% is obtained from XRD upon full delithiation, while the corresponding XAS-derived value yields a slightly smaller contraction of 5.0%. Similarly, $LiCo_{0.5}Mn_{1.5}O_4$ exhibits volume changes of 4.4% and 3.5% from XRD and XAS, respectively, while $LiCoMnO_4$ shows the smallest contractions of 2.1% and 1.6% from XRD and XAS, respectively. A systematic trend is thus observed whereby increasing Co substitution progressively decrease the unit cell volume change during charge, consistent with the greater structural rigidity induced by the more covalent Co-O bonds, in agreement with EXAFS fitting results (**section 2.4.2**).

While both XRD- and EXAFS-derived structural parameters exhibit a similar trend in volume variation during cycling, the XRD-derived unit-cell volume shows higher reversibility, as the volume after discharge returns closer to the pristine state than the local structural volume evolution reconstructed from EXAFS. This difference highlights the distinct structural sensitivities of the two techniques. XRD probes the long-range average crystal structure and determines the average unit-cell parameters based on the periodic arrangement of atoms throughout the bulk. Consequently, if most transition-metal sites recover their original coordination environment upon re-lithiation, the Bragg-derived lattice parameters can return close to their initial values, even when a fraction of sites remains locally distorted. In contrast, EXAFS is an element-specific probe of the local coordination environment around the absorbing atom and is sensitive to short-range structural variations within only a few coordination shells. It can therefore detect subtle changes in TM-O and TM-TM bond distances, local strain, and static disorder that may be averaged out in diffraction measurements. [26],[27] Reported XRD and EXAFS investigations of defective $LiMn_2O_4$ spinels and Co-substituted species have demonstrated that local $MnO_6$ octahedral distortions, associated with Jahn-Teller active $Mn^{3+}$ species and oxygen/cation defects introduced during synthesis or electrochemical cycling, can persist locally without producing significant changes in the average lattice parameters. [18],[28],[29] Therefore, the discrepancy between the two techniques indicates that the recovery of the average crystallographic lattice does not necessarily correspond to a complete relaxation of the local transition-metal environment. The remaining local distortions may originate from incomplete relaxation of the $MO_6$ octahedra, rearrangement of the oxygen sublattice, or defect-induced structural disorder generated during high-voltage cycling.

The progressive reduction in unit cell volume change with increasing Co substitution, particularly when compared to the pure Mn spinel $LiMn_2O_4$, highlights the significant role of ionic radii in governing the structural rigidity of the spinel framework. This is supported by the EXAFS analysis of the reconstructed principal components across all three compositions ($LiMn_2O_4$, $LiCo_{0.5}Mn_{1.5}O_4$, and $LiCoMnO_4$) for the *Pristine*, *Intermediate*, and *Final Charged* species. Highly Co-substituted compositions exhibit smaller ionic radii variations as the cell undergoes electrochemical cycling, reflecting a stiffer local coordination environment. This structural rigidity is further supported by the Debye-Waller factors obtained from the EXAFS analysis (**Table S2-S4**), which show minimal variation across different states of charge for Co-substituted materials. These results suggest a more covalent character of the Co-O bond relative to the Mn-O bond within the spinel lattice. The reduction in volume variation with increasing Co content is consistent with the broader tendency of Co-rich oxide cathodes to exhibit comparatively small structural changes upon cycling, in contrast to Mn-rich systems that generally undergo larger lattice variations. [30]-[31] However, the greater structural rigidity of highly Co-substituted compositions also appears to be associated with increased capacity irreversibility relative to $LiMn_2O_4$, possibly because of the reduced structural flexibility needed to accommodate full lithium reinsertion upon discharge.

Beyond minimizing volume changes, increasing Co substitution progressively raises the working potential of the material, as evidenced by the XANES spectra the Mn and Co K-edges across the three compositions. The $Mn^{4+}/Mn^{3+}$ redox couple is active exclusively within the lower voltage regime (below ~4.5 V), while the $Co^{4+}/Co^{3+}$ redox couple is activated only at higher voltages (above ~4.7 V), enabling high-voltage electrochemical performance. The *operando* XAS measurements employed in this study allow this tandem redox behavior of Mn and Co to be directly and unambiguously resolved, a level of mechanistic detail that is not accessible through *operando* XRD alone.

## 2. Conclusion

Understanding the composition-dependent volume evolution of spinel cathodes is essential for mitigating mechanical degradation and improving the cycling stability of high-voltage lithium-ion batteries. In this study, a correlative *operando* XRD and XAS approach was applied to the $LiCo_xMn_{2-x}O_4$ spinel system to disentangle the contributions of Mn and Co redox processes to the structural evolution during cycling.

Electrochemical and *operando* XAS measurements confirm a clear separation of the $Mn^{4+}/Mn^{3+}$ and $Co^{4+}/Co^{3+}$ redox processes, with Mn redox dominating below ~4.7 V and Co redox above this potential. Combining XRD and XAS further enabled the evolution of the different structural states to be tracked throughout cycling. Despite probing different structural length scales, both techniques yielded closely consistent volume-evolution profiles, demonstrating the robustness of the correlative approach.

A clear composition-dependent trend was observed: increasing Co substitution progressively suppresses the volume change during cycling, from 5.0% to 1.6% based on XAS and from 5.8% to 2.1% based on XRD. These results highlight the effectiveness of Co substitution in reducing electrochemically induced mechanical strain in spinel cathodes.

More broadly, this correlative framework provides a valuable strategy for linking cation-specific redox activity to structural and volumetric evolution and could be extended to the design and screening of zero-strain or near-zero-strain electrode materials for high-energy, long-life lithium-ion batteries.

# 3. Methods

## 3.1. Synthesis

$Li_2CO_3$ (Sigma-Aldrich, ≥99%), $Co_3O_4$ (Thermo Scientific, 99.7%) and $MnO_2$ (Thermo Scientific, 99.9%) were used as precursors for the solid-state synthesis of $LiMn_2O_4$, $LiCo_{0.25}Mn_{1.75}O_4$, $LiCo_{0.5}Mn_{1.5}O_4$, $LiCo_{0.75}Mn_{1.25}O_4$ and $LiCoMnO_4$. The precursors were mixed in stoichiometric amounts and ground using an agate pestle and mortar, pressed into pellets under 43 MPa and placed in an alumina crucible for two successive heat treatments. The sintering was done in a muffle furnace under atmospheric air condition.

The pellets were first heated at 800 °C for 12 hours with a heating rate of 300 °C/h. After that the pellets were cooled down to room temperature with a cooling rate of 1 °C/h. The samples were ground and pelleted again. A second heat treatment at 600 °C for 72 hours using the same heating and cooling rates was performed. The obtained samples were finally stored and used in a glovebox maintained under Argon atmosphere.

## 3.2. X-Ray Diffraction (XRD)

Powder X-Ray Diffraction were carried out on a Bruker D8 Advance diffractometer in Bragg-Brentano θ-θ geometry, equipped with a LynxEye detector and $CuK_\alpha$ radiation ($\lambda\ K_{\alpha1}$ = 1.54056 Å, $\lambda\ K_{\alpha2}$ = 1.54439 Å). Powder diffraction patterns for pristine samples were collected in the range 2θ = (10-80) °, with 0.01 ° steps and 0.5 s acquisition. Rietveld refinements were done using Fullprof. [32]

## 3.3. Electrochemistry

Positive electrodes were prepared as self-standing electrodes made of active material, polytetrafluoroethylene (PTFE) and carbon black (ratio 55/15/30), rolled and folded multiple times to ensure homogeneity. The mass loading of active material was optimized for the XAS measurement, resulting in 10 mm diameter electrodes with active material loadings between 5 and 7 mg/cm$^2$. Coin cells were assembled using an Al current collector on the cathode side, 1 layer of glass fiber separator (Whatman GF/C type), metallic Li as an anode and 1 M $LiPF_6$ dissolved in a 1:1 volume ratio of ethylene carbonate (EC) and dimethyl carbonate (DMC) as electrolyte (LP30). The cells were sealed under Ar in a glovebox and cycled using a Biologic VMP3 potentiostat. For the preparation of ex situ samples, the cycled coin cells were stopped at selected states of charge and were opened in the glovebox, the self-standing electrode was recovered, washed three times in DMC to rinse off salts, and dried under vacuum before storage.

Note that an increased polarization is expected for the electrochemical profiles obtained from the *operando* cell relative to conventional coin cells, as a consequence of the *operando* cell geometry, which does not affect the redox assignments or voltage windows discussed in this work.

### 3.4. *Operando* XRD

*Operando* XRD was performed on the same diffractometer as for powder samples, using a Leriche cell in reflection [33] equipped with a Be window and a 3 µm thick Al film as current collector on the cathode side. Two layers of glass fiber separator (Whatman GF/C type) were used, with metallic lithium as an anode and LP30 electrolyte. A key challenge in *operando* XRD is acquiring sufficient crystallographic information within a sufficiently short acquisition time to accurately capture the continuous structural changes occurring during electrochemical cycling. Hence, the diffraction measurements were restricted to a narrow 2θ range (36 - 39°), encompassing the 311 and 222 reflections, which are sensitive to changes in the lattice parameter. 10-min scans were measured continuously on this 2θ range at a relatively fast cycling rate (C/5). The data was first processed sequentially using the Fullprof App software [34] and the result of the initial refinement was improved manually afterwards to extract lattice parameter *a*. Slower acquisition with a larger angular range were also done to make sure that no other structural evolution than the delithiation/lithiation of the materials studied were observed.

### 3.5. *Operando* XAS

*Operando* X-ray absorption spectroscopy (XAS) measurements were performed at the ROCK beamline [35] of the SOLEIL synchrotron (France) at both the Mn and Co K-edges. The XAS spectra were collected with an acquisition time of 2 min per spectrum using a Si(111) quick-XAS monochromator operating at an oscillation frequency of 2 Hz and a beam size of 2.35 × 0.5 mm. Energy calibration and spectral normalization were carried out using the python graphical interfaces available at the ROCK beamline. A 100 µm-thick Al filter was placed before the I0 monitor to attenuate the incident beam.

The beamline is equipped with a dedicated sample holder that enables simultaneous measurements of up to three electrochemical cells. *Operando* experiments were conducted using the Leriche cell for transmission mode with X-ray transparent beryllium windows. [36] Three cathode active materials were investigated: $LiMn_2O_4$, $LiCo_{0.5}Mn_{1.5}O_4$, and $LiCoMnO_4$. The electrochemical cycling conditions, including electrode mass loading, cycling rate, and voltage window, are summarized in **Table S2**. Self-standing electrodes of 10 mm diameter were cycled against metallic Li, with a 3 µm-thick Al film as current collector on the cathode side, one separator (GF/C type), and 3 discs of Li metal on the anode side, using LP30 electrolyte. To evaluate potential beam-induced damage, measurements were collected at two distinct positions at both the open-circuit voltage (OCV) and the end of charge. The resulting XANES spectra (**Figure S7**) show no discernible differences between the two positions at either state of charge, confirming that beam-induced effects on the sample are negligible under the employed measurement conditions.

In addition, several *ex situ* samples of the $LiCo_xMn_{2-x}O_4$ series (x = 0, 0.25, 0.5, 0.75, and 1) were analyzed. Reference compounds for Mn ($LiNi_{0.5}Mn_{1.5}O_4$, $Li_2MnO_3$, $LiMnO_2$) and Co ($Co_3O_4$, CoOOH, $LiCoO_2$, NMC111 pristine and charged state) were also measured to assess the Mn and Co oxidation states in $LiCo_xMn_{2-x}O_4$, respectively.

### 3.6. XANES and EXAFS analysis

The quick-XAS datasets, arranged as matrices of consecutive spectra after energy calibration and spectra normalization, were analyzed using a combination of Principal Component Analysis (PCA) and Multivariate Curve Resolution with Alternating Least Squares (MCR-ALS), following the methodology previously reported in the literature.[37] PCA was first employed to determine the number of statistically significant components contributing to the spectral evolution. Initial estimatation of the pure component spectra were then obtained using an algorithm that identifies the most different spectra within the dataset (i.e., the purest spectra). MCR-ALS was subsequently applied to extract the concentration profiles and reconstructed component spectra, with non-negativity constraints imposed on both the concentration and spectral profiles throughout the optimization.

The EXAFS oscillations of the reconstructed components by MCR-ALS, together with that of the pristine sample, were analyzed in R space using the Artemis interfaces of the Demeter software package [38]. Theoretical scattering paths were calculated using FEFF based on the Fd-3m $LiMn_2O_4$ spinel structure. To reflect the actual transition metal composition of each sample, the Mn site in the FEFF input was partially substituted with Co according to the nominal Co:Mn occupancy (x = 0, 0.5, 1.0), ensuring that both the absorbing atom (Mn or Co, depending on the edge probed) and the surrounding TM shell composition matched the sample stoichiometry. Co was placed on the octahedral (16d) site. XRD rietveld refinement showed partial Co occupancy of the tetrahedral (8a) site for $LiCoMnO_4$ (~13% of Co), but including this site in the EXAFS model, with the fraction fixed to the XRD value, degraded the fit (R-factor 0.0018 → 0.0157). This lower fit might suggest an overestimated Co content value from XRD due to the similarity of X-ray scattering factors of Co and Mn. The octahedral-only model was therefore used. The $k^2$-weighted EXAFS spectra were Fourier transformed over a k range of 3.0 – 9.9 $Å^{-1}$ for Mn K-edge (due to the presence of Fe in Be window), and 2.6 – 11.6 $Å^{-1}$ for Co K-edge, using a sine window function. The fitting was subsequently performed in R space over the range of 0-3.8 Å ($\Delta R$ = 3.8 Å) using a sine window function. The coordination number (N) of the TM-O first-shell were fixed at 6 for all compositions. For $LiMn_2O_4$, the N values for the second-shell Mn-Mn were fixed at 6. For $LiCo_{0.5}Mn_{1.5}O_4$, the N values of the second-shell Mn-Mn and Mn-Co were fixed at 4 and 2, respectively. For $LiCoMnO_4$, the N values for the second-shell Co-Co and Co-Mn were fixed at 3 and 3, respectively. The amplitude reduction factors ($S_0^2$) were fixed at 0.75 for both Mn and Co K-edges.

The interatomic distances (R) of all scattering paths were allowed to vary during the refinement. The energy shift parameter ($\Delta E_0$) was initially refined on the pristine material and subsequently fixed to the optimized value for the final fitting procedure. The Debye-Waller factors ($\sigma^2$) associated with the first-shell TM-O paths were refined, whereas those corresponding to the TM-TM scattering paths (Mn-Mn, Mn-Co, Co-Co, and Co-Mn) were fixed because of the limited number of independent parameters.

## Supplementary information

Supplementary data to this article can be found online at

Details on…

## Notes

The authors declare no competing financial interest.

## Acknowledgements

This work benefited from state support managed by the National Research Agency (ANR) for France 2030 under the reference ANR-22-PEBA-0004 (PEPR- LIMASSE project) and ANR-22-PEBA-0005 (PEPR OPENSTORM). The authors acknowledge the French Network on Electrochemical Energy Storage (RS2E) for its support through the Labex STORE-EX (ANR10LABX-76-01). The XAS experiment was performed at the ROCK beamline (public grant overseen by the French National Research Agency (ANR) as part of the "Investissements d'Avenir" program reference: ANR-10-EQPX-45) of SOLEIL synchrotron. The authors acknowledge the technical staff of Laboratoire de Chimie de la Matière Condensée de Paris (LCMCP) and Physicochimie des Electrolytes et Nanosystèmes interfaciaux (PHENIX). The authors thank the RMES team (Laboratoire de Chimie de la Matière Condensée de Paris - LCMCP) for its help and support.

# Supplementary Information

## Insights on the volume variation of $LiCo_xMn_{2-x}O_4$ electrode upon cycling: a correlative *operando* XRD and XAS study

Eunike Mahayoni[1], Louiza Larbi[2], Romain Batiot[2], Anatole Weill[2], Damien Brégiroux[2], Antonella Iadecola[1,3], Arnaud Perez[2,3,*]

[1] *Sorbonne Université, CNRS, Physicochimie des Electrolytes et Nanosystèmes interfaciaux (PHENIX), F-75005, Paris, France*

[2] *Sorbonne Université, CNRS, Laboratoire de Chimie de la Matière Condensée de Paris (LCMCP), F-75005, Paris, France*

[3] *Réseau sur le Stockage Electrochimique de l'Energie (RS2E), FR CNRS 3459, F-80039, Amiens, France*

[*] Corresponding author: arnaud.perez@sorbonne-universite.fr

**List of tables**

**List of figures**

# 1.S1. Extended data

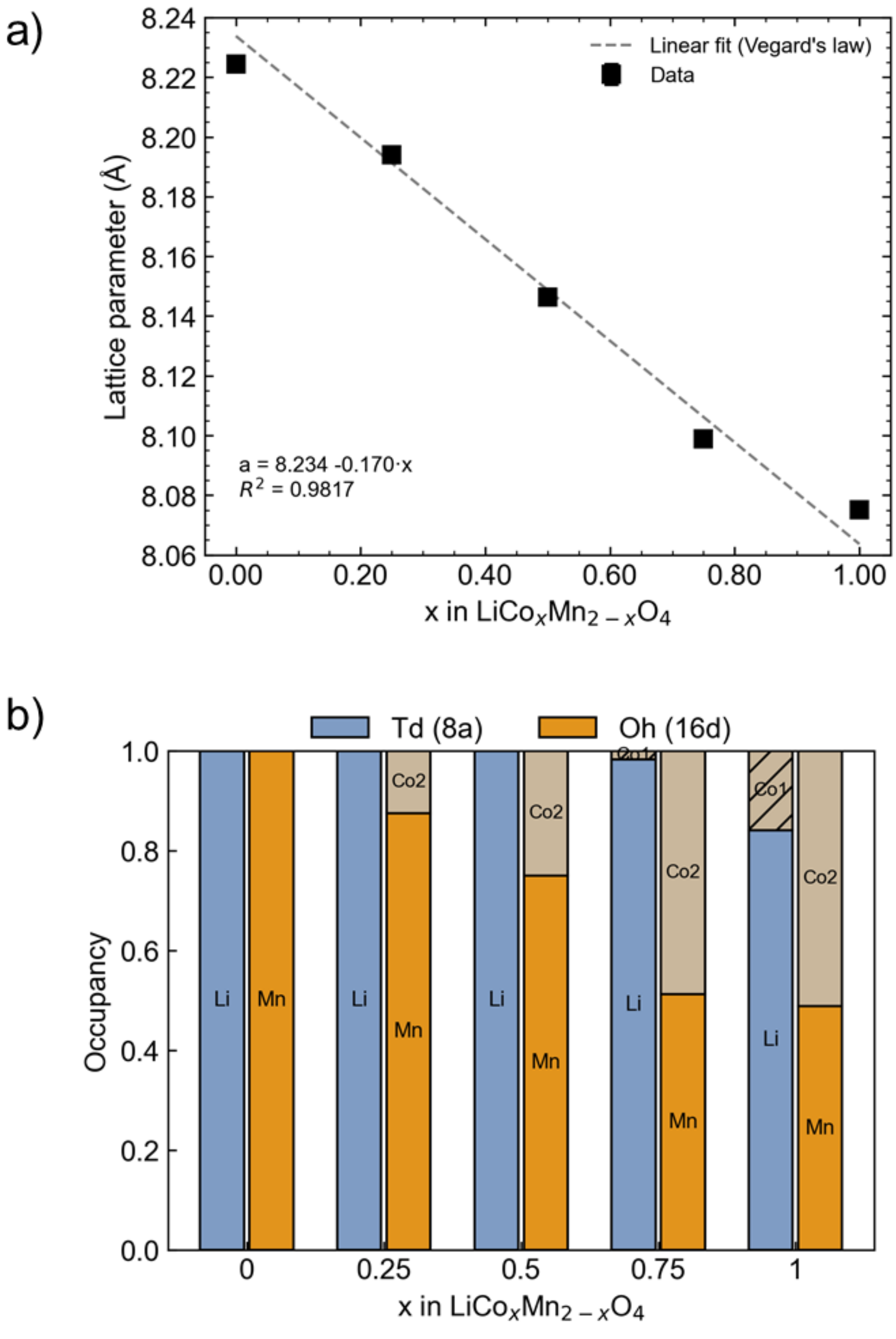

**Figure S2 |** a) Lattice parameters of the $LiCo_xMn_{2-x}O_4$ materials and its linear fit (Vegard's law). b) Occupancy of thetrahedral (Td) and octahedral (Oh) sites by Li, Mn and Co cations in $LiCo_xMn_{2-x}O_4$ spinel

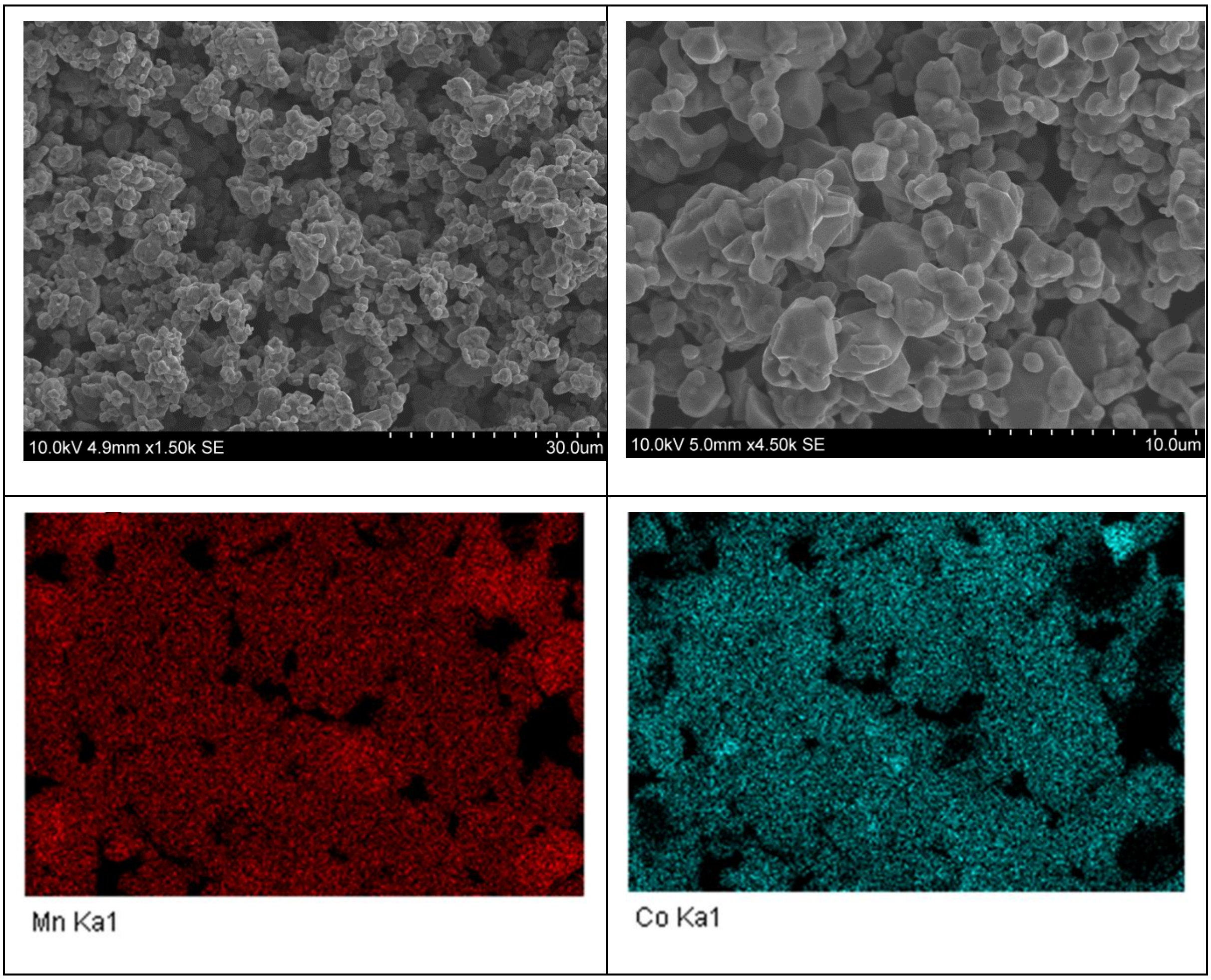

**Figure S3** | SEM and EDX images of $LiCoMnO_4$ sample

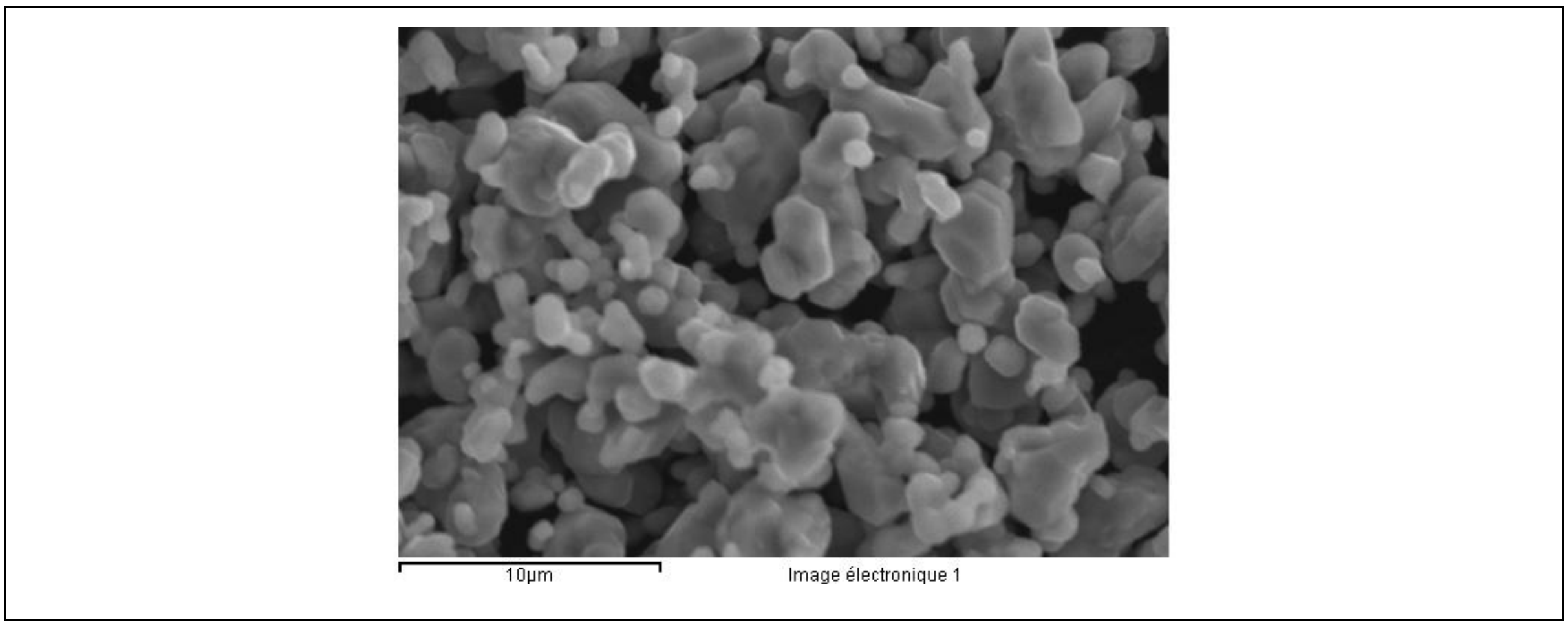

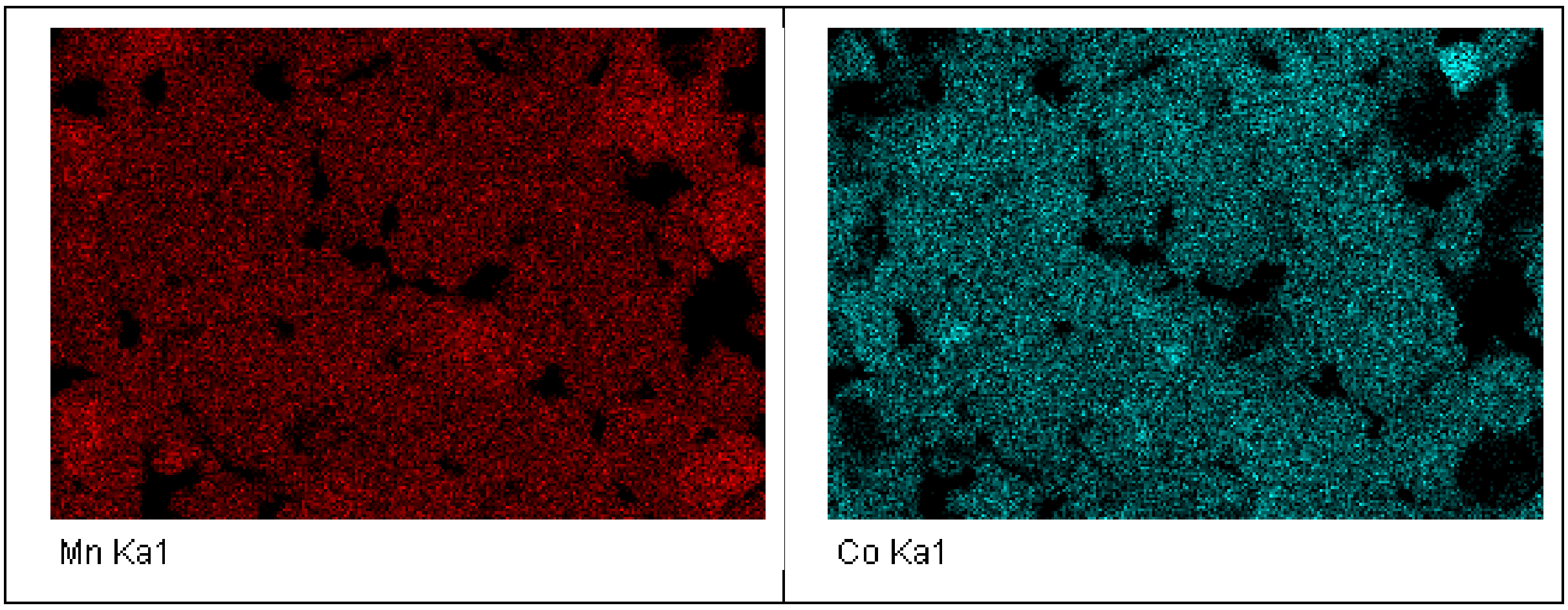
Mn Ka1
Co Ka1

**Table S1** | Details of electrochemical cells measured during *operando* XRD and XAS measurement.

| Cathode active material | ***Operando XRD*** | | | ***Operando XAS*** | | |
|---|---|---|---|---|---|---|
| | **Mass loading** | **Cycling rate** | **Voltage Window** | **Mass loading** | **Cycling rate** | **Voltage Window** |
| $LiMn_2O_4$ | 3.98 mg (56.1% active material) | C/5 (150 μA cm$^{-2}$) | 3 V – 4.5 V (1st cycle)<br>3 V – 5.3 V (2nd cycle) | 8.3 mg (56.1% active material) | C/5 (176 μA cm$^{-2}$) | 3 V – 4.5 V (2 cycles) |
| $LiCo_{0.5}Mn_{1.5}O_4$ | 6.3 mg (56.5% active material) | C/5 (228 μA cm$^{-2}$) | 3 V – 4.5 V (1st cycle)<br>3 V – 5.3 V (2nd cycle) | 6.3 mg (57.5% active material) | C/2 (339 μA cm$^{-2}$) | 3 V – 4.5 V (1st cycle)<br>3 V – 5.4 V (2nd cycle) |
| | | | | 5.9 mg (57.5% active material) | C/5 (219 μA cm$^{-2}$) | 3 V – 4.5 V (1st cycle)<br>3 V – 5.1 V (2nd cycle) |
| $LiCoMnO_4$ | 2.77 mg (56.0% active material) | C/5 (102 μA cm$^{-2}$) | 3 V – 5.3 V (2 cycles) | 6.4 mg (56.0% active material) | C/5 (133 μA cm$^{-2}$) | 3 V – 5.3 V (1st cycle) |

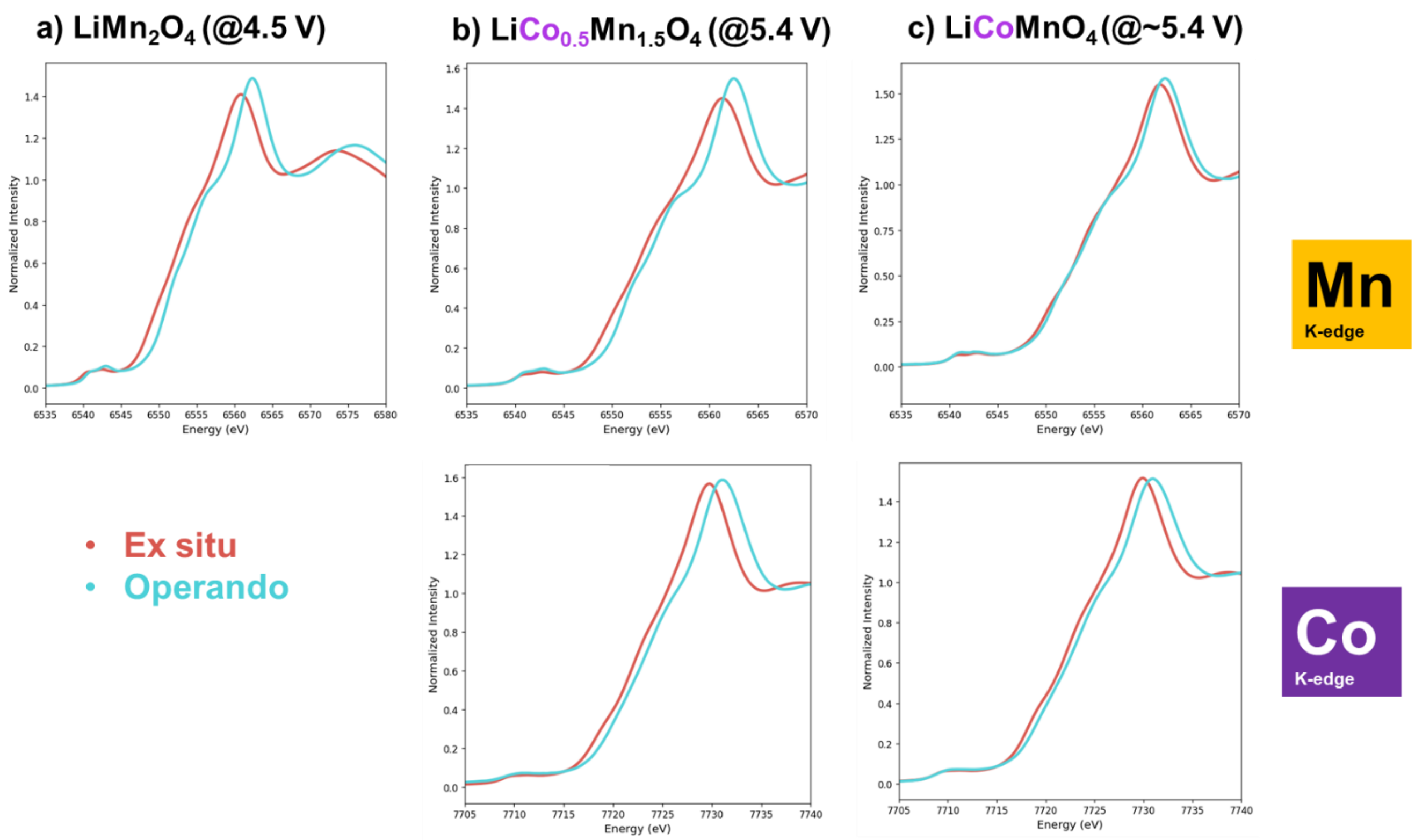

**Figure S4** | Comparison of XANES spectra on *ex situ* electrodes and the corresponding end of charge obtained in *operando* conditions.

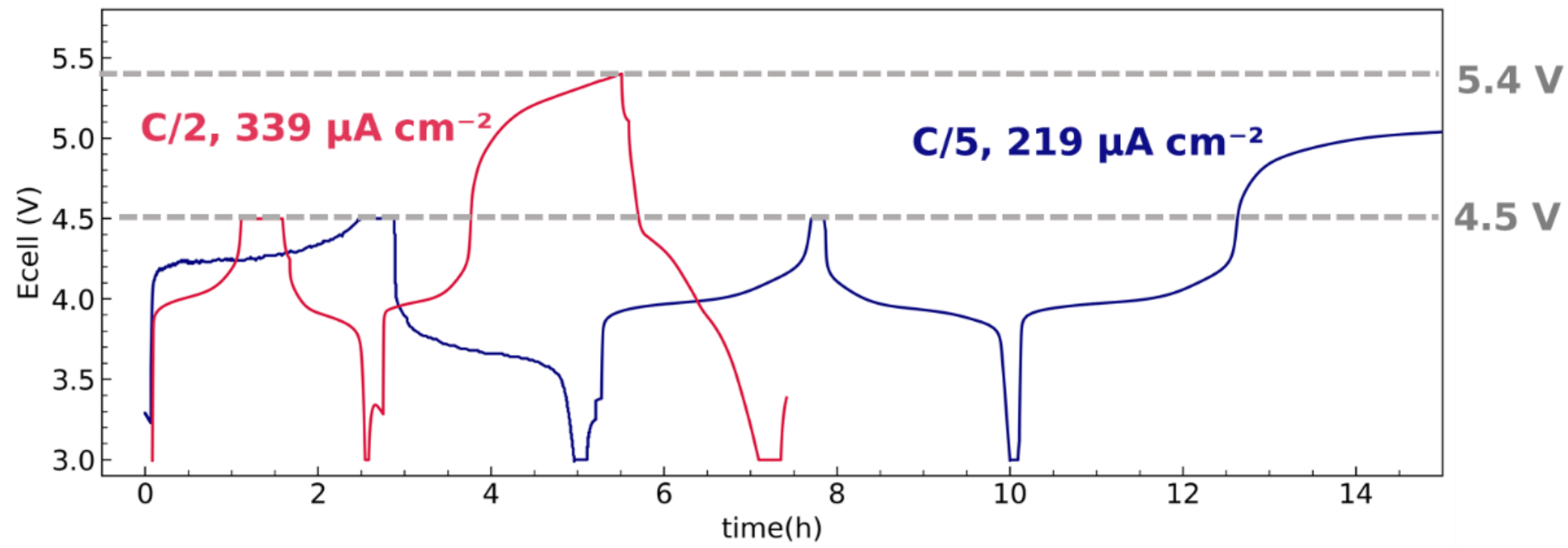

**Figure S5** | GCPL profile of two different $LiCo_{0.5}Mn_{1.5}O_4$ cell at C/2 (red) and C/5 (blue) rates. Only the cell with the faster rate (C/2, red) could reach higher voltage of 5.4 V, meanwhile the cell cycled at C/5 only reached 5.1 V vs $Li^+$/Li.

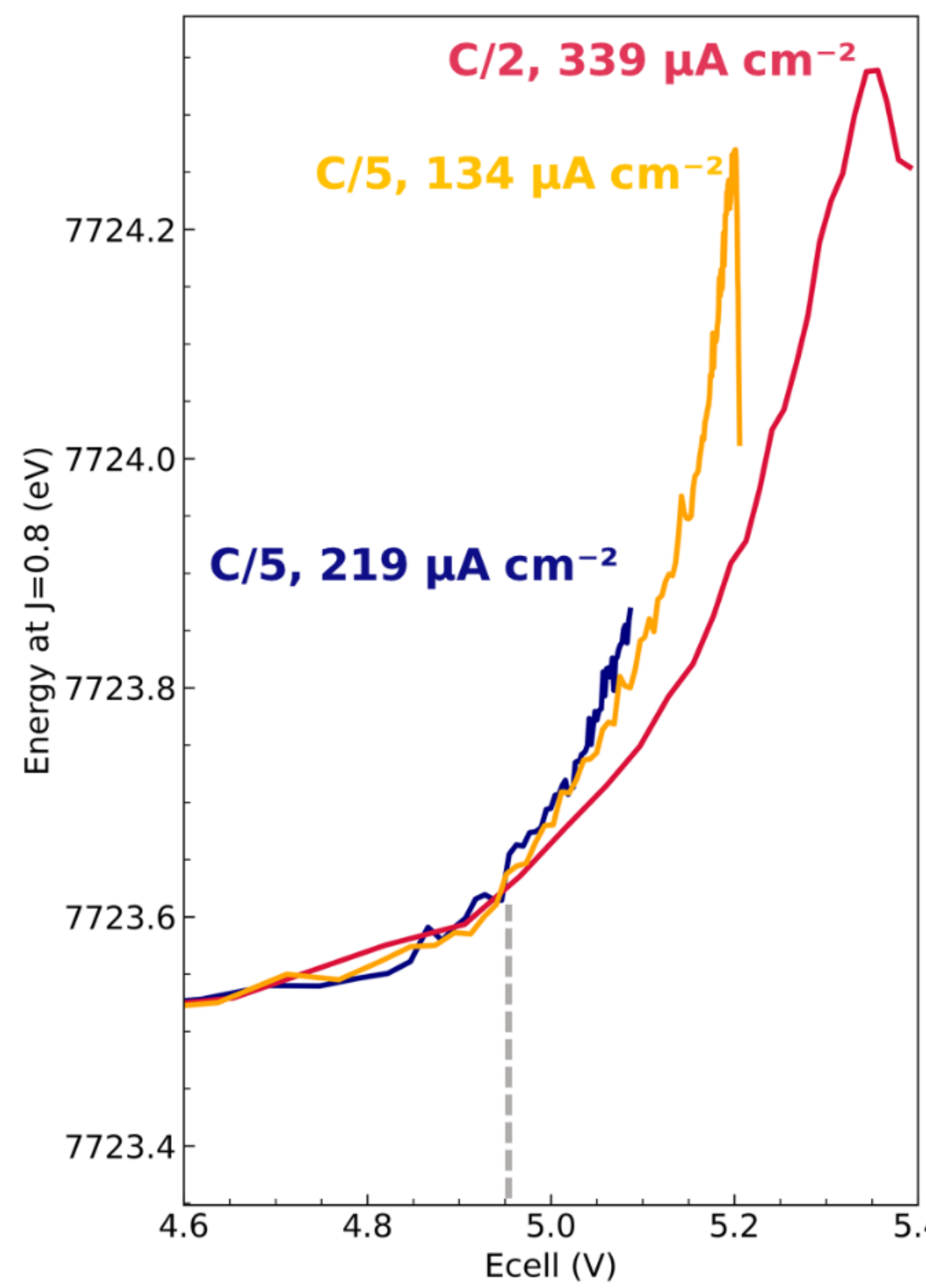

**Figure S6** | Edge position of Co K-edge XANES spectra considering the normalization jump J=0.8 as a function of potential for $LiCo_{0.5}Mn_{1.5}O_4$ cells at different rates, C/2 (red), and C/5 (blue and yellow). Below 4.9 V, the reaction is not affected by the cycling rate. Above 4.9 V, only C/2 faster rate enables to reach 5.4, revealing that this rate is required to drive the electrochemical reaction above 5.1 V. This behavior is consistent with a

competition between the delithiation reaction driven by the applied current and re-intercalation associated with electrolyte oxidation at high potential, as already demonstrated for K-ion batteries [23]. At C/5, the slow kinetics enhance parasitic electrolyte degradation, whereas the faster C/2 rate reduces the extent of this competing process and allows the cell to access the higher-voltage Co redox region.

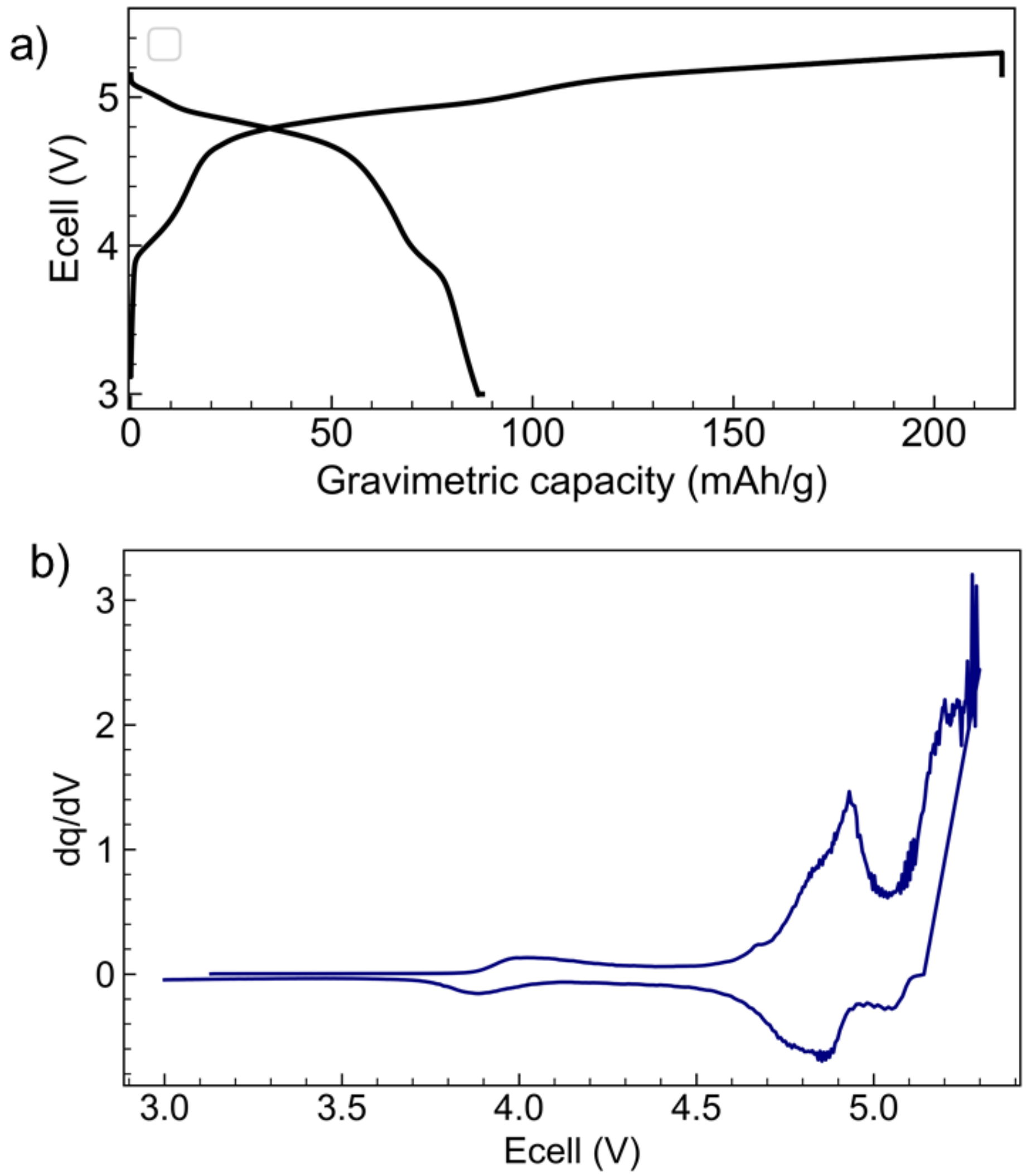

**Figure S7** | a) GCPL profile for LiCoMnO4 XAS *operando* cell charged at C/5 (133 µA $cm^{-2}$) and its b) derivative cure. The charge capacity reaches approximately 220 mAh/g, exceeding the theoretical capacity of ~145 mAh/g for full delithiation of LiCoMnO4. This excess capacity indicates that a fraction of the charge passed above ~5.0 V is possibly being consumed by a parasitic process rather than Li extraction. This is supported by the dq/dV profile (b) which shows a continuously rising, non-decaying signal above ~5.0 V rather than a discrete redox peak, that could be a sign of irreversible oxidation process. Hence, the voltage limit was set at 5.3 V instead of 5.4 V to preserve coulombic efficiency.

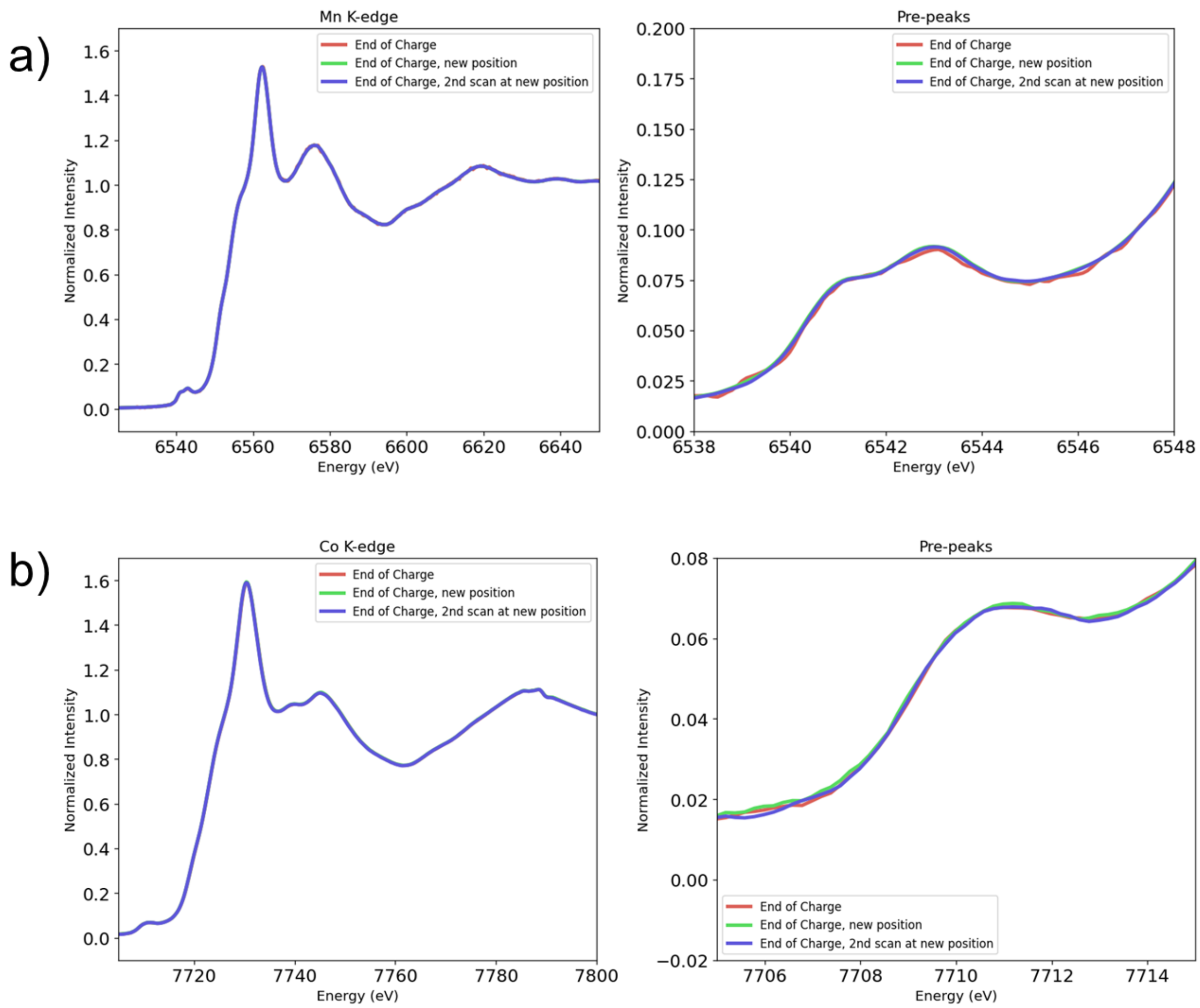

**Figure S8** | Comparison of the $LiCo_{0.5}Mn_{1.5}O_4$ electrode at high voltage end of charge (5.4 V) on both Mn K-edge (a) and Co K-edge (b), by doing two scans on a different vertical position. No significative differences are observed between the spectra, confirming the absence of any beam effect during the operando XAS experiment.

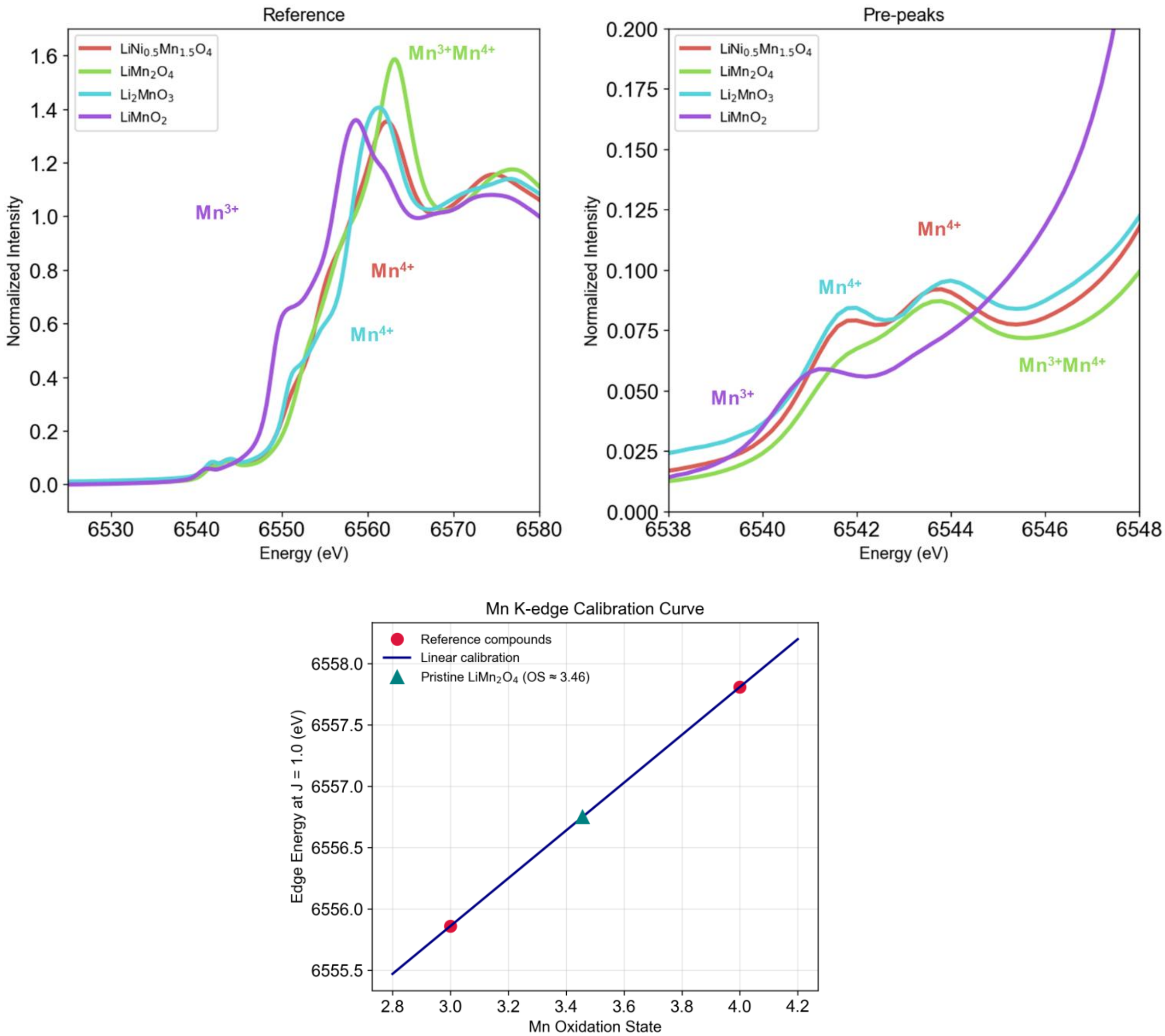

**Figure S9** | Top: Mn-based materials used as references for $LiCo_xMn_{2-x}O_4$ along with a zoom of the pre-peak region (right panel). Down: Linear calibration curve to assess on the Mn oxidation state of $LiMn_2O_4$.

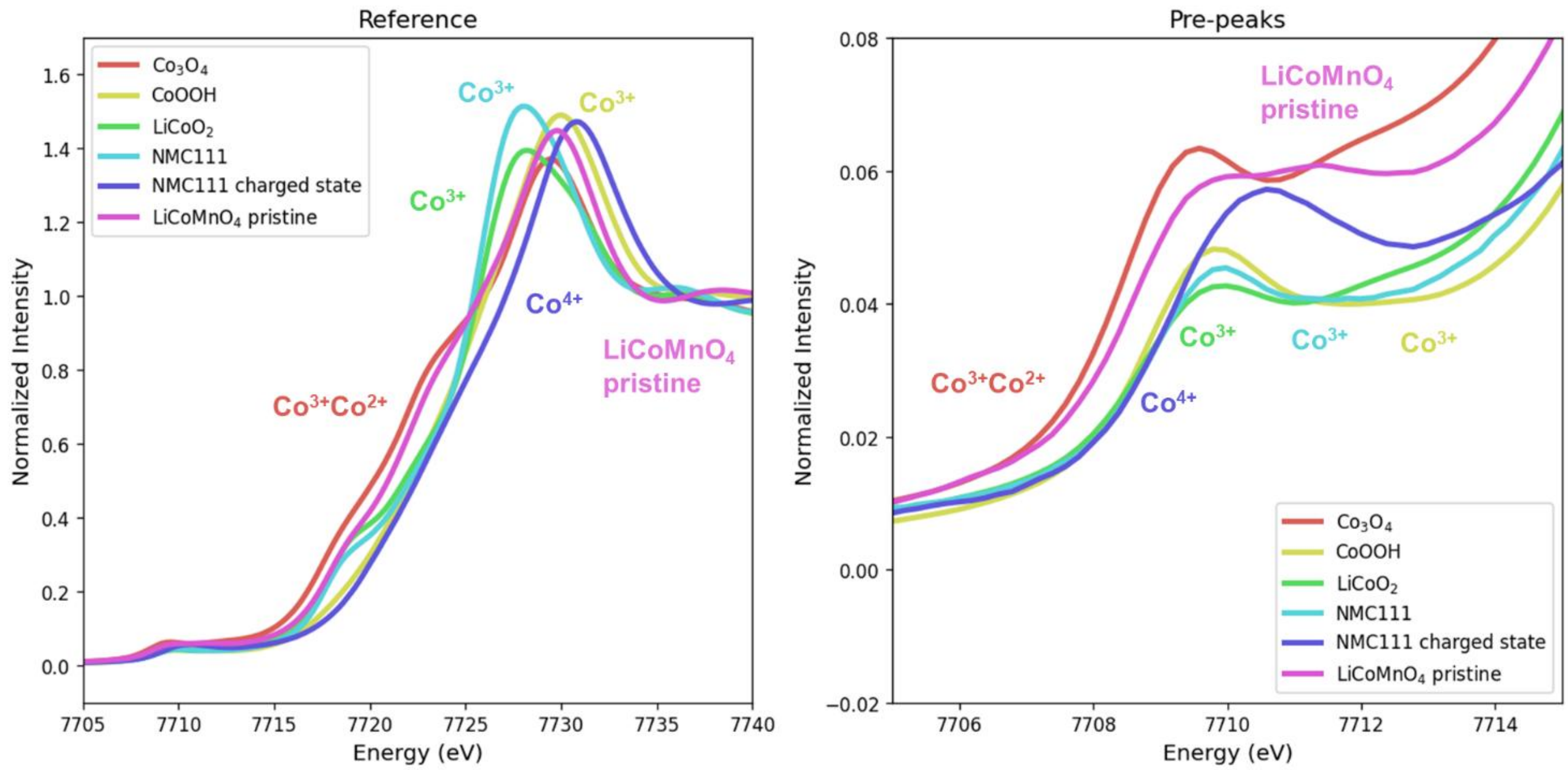

**Figure S10** | Co-based materials used as references for $LiCo_xMn_{2-x}O_4$. A zoom of the pre-peak region is reported in the right panel.

# 2.S2. Supplementary Analysis

## 2.1. S2.1 MCR-ALS

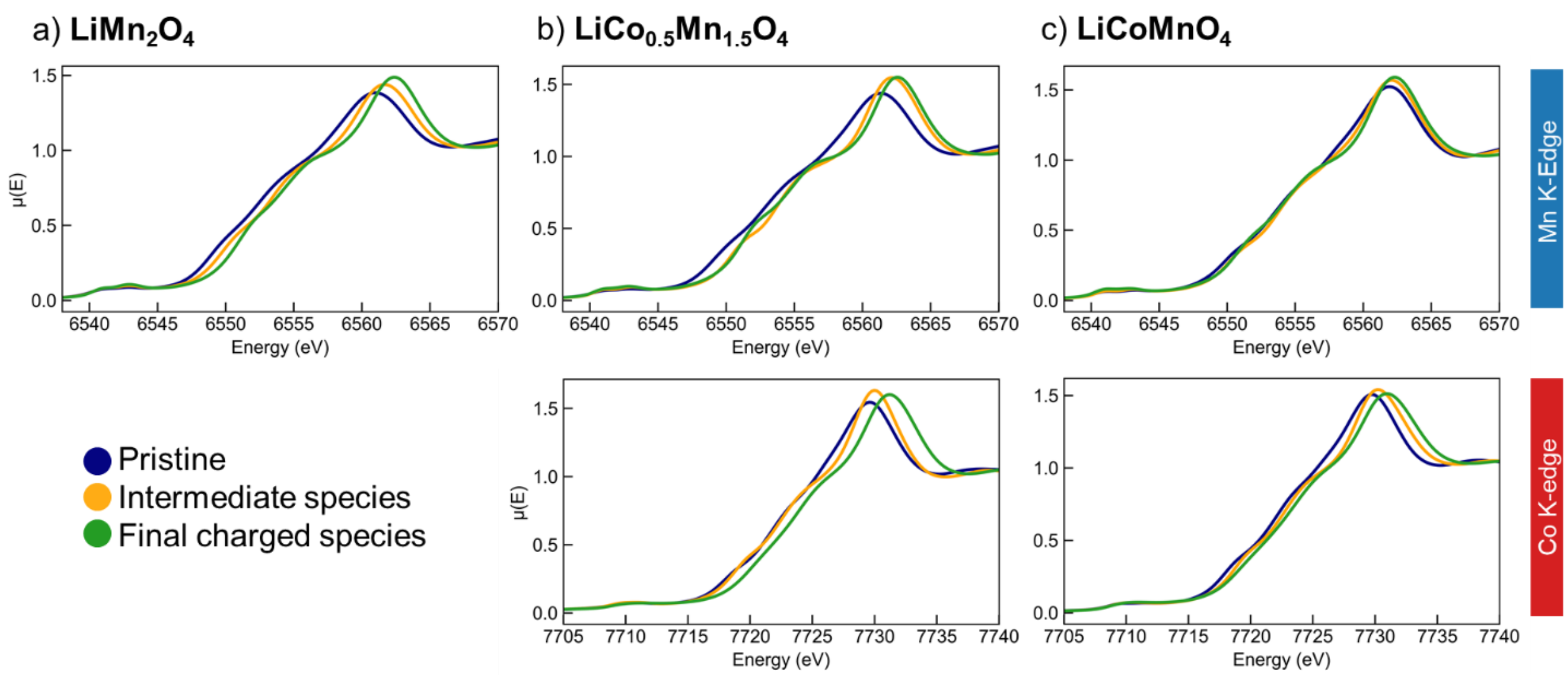

**Figure S11** | XANES spectra of reconstructed principal components from PCA and MCR-ALS analysis on the $LiCo_xMn_{2-x}O_4$ samples, with three different pure components: pristine (blue), intermediate (yellow), and final charged (green) species, on the Mn K-edge (top row) and Co K-edge (bottom row), for a) $LiMn_2O_4$, b) $LiCo_{0.5}Mn_{1.5}O_4$, and c) $LiCoMnO_4$.

The reconstructed principal component XANES spectra of $LiMn_2O_4$ charged to 4.5 V (**Figure S11 a**) exhibit a gradual shift toward higher energies at the Mn K-edge, indicative of progressive Mn oxidation throughout the charging process. The corresponding concentration profiles (**Figure S12 a**) reveal that the Pristine species is gradually consumed during charge, giving rise to the Intermediate species at mid-charge, before ultimately yielding the Final Charged species as the sole observed phase at the end of charge. A similar evolution is observed during discharge, indicating reasonably good electrochemical reversibility. However, at the end of the first discharge, a residual fraction of the Final Charged species (5.6%) persists alongside the re-emerging Pristine species, suggesting incomplete reduction. During the second charge, the Pristine species is once again consumed, with the Final Charged species becoming the only observed phase by the end of charge. At the end of the second discharge, a comparable residual fraction of the Final Charged species (4.9%) is retained, consistent with the behavior observed at the end of the first discharge cycle. Overall, the charge-discharge process proceeds in a largely reversible and symmetrical manner across both cycles, with the minor retention of the Final Charged species at the end of each discharge representing the only notable deviation from complete reversibility.

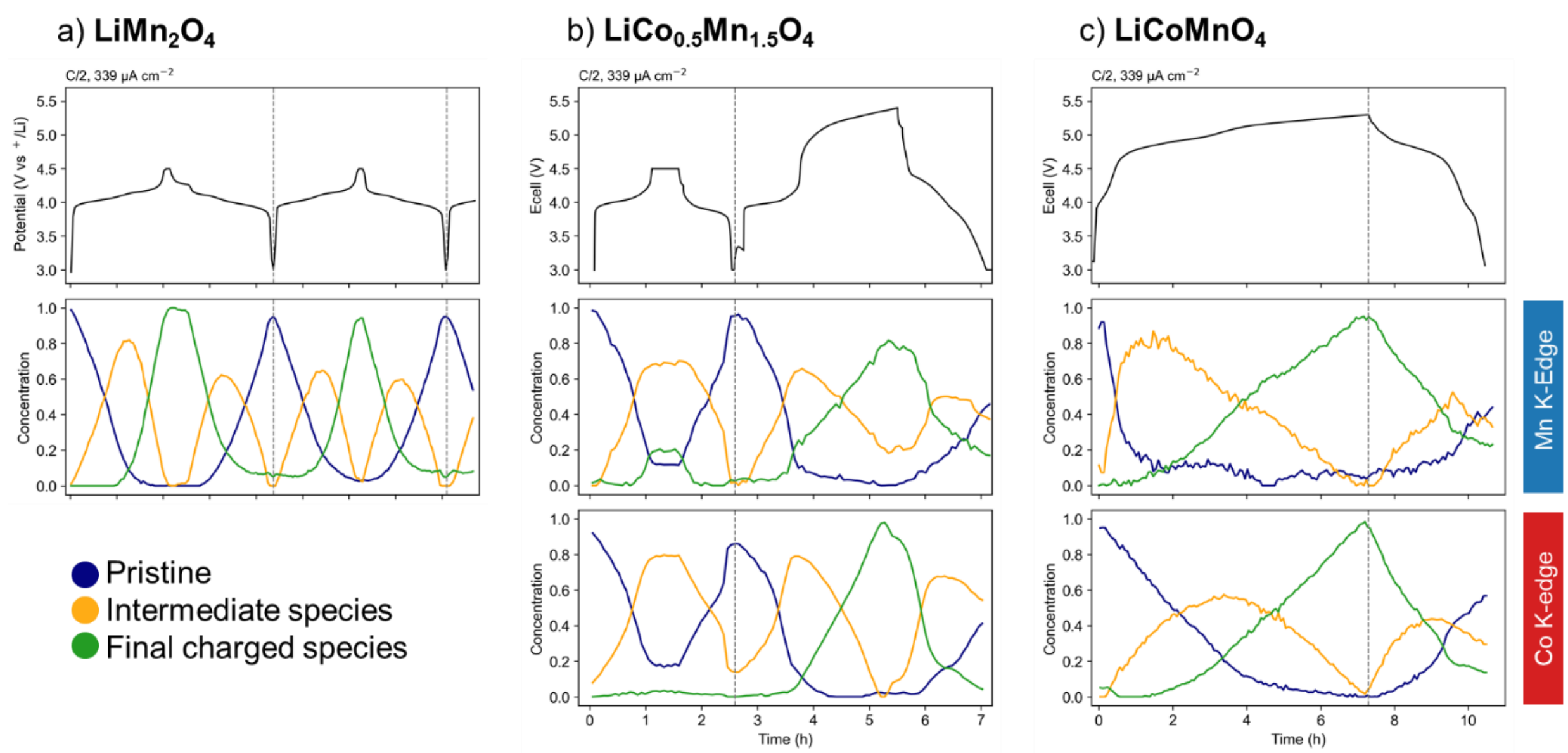

**Figure S12** | Concentration profile of reconstructed principal components from PCA and MCR-ALS analysis on the $LiCo_xMn_{2-x}O_4$ samples, with three different pure components: pristine (blue), intermediate (yellow), and final charged (green) species, on the Mn K-edge (middle row) and Co K-edge (bottom row), for a) $LiMn_2O_4$, b) $LiCo_{0.5}Mn_{1.5}O_4$, and c) $LiCoMnO_4$, with corresponding GCPL profile of potential as a function of time (top row) for each component.

For $LiCo_{0.5}Mn_{1.5}O_4$, the Mn K-edge XANES spectra (**Figure S11 b, top**) of the Intermediate and Final Charged species are notably similar: the Mn centre is already largely oxidized at the Intermediate stage, with only marginal additional oxidation occurring upon reaching the Final Charged state. Looking at the corresponding concentration profile at the Mn K-edge (**Figure S12 b, top**), the first charge, limited at 4.5 V, yields a state at the end of charge where all three species coexist, with the Intermediate species predominating at the highest concentration (~80%). This indicates that the Intermediate species in this composition is primarily associated with Mn oxidation occurring within the lower voltage regime. Upon subsequent discharge, the Pristine species is largely recovered. During the second charge, in which the system is driven to a higher cutoff voltage of 5.4 V, the Intermediate species emerges rapidly during the lower voltage region before its concentration diminishes as the Final Charged species grows in, ultimately becoming the most concentrated species at the end of charge. The appearance of the Final Charged species at higher voltages is attributed to changes in the Mn redox environment induced by the oxidation of the neighbouring Co centres.

At the Co K-edge (**Figure S11 b, bottom**), the XANES spectrum of the Intermediate species closely resembles that of the Pristine species, while the Final Charged species exhibits a clear shift toward higher energies, consistent with a more oxidized Co centre. The corresponding concentration profile **Figure S12 b, bottom**) supports this interpretation: the Intermediate species reaches its maximum concentration at the end of the first charge at 4.5 V, likely reflecting the indirect influence of Mn oxidation on the local Co environment, consistent with the behaviour observed at the Mn K-edge. The Final Charged species, by contrast, reaches its maximum concentration at the end of the second charge at 5.4 V, consistent with Co oxidation occurring exclusively at higher voltages. The close correspondence between the Mn and Co K-edge concentration profiles underscores the strong interdependence of the two redox centres, whereby the local structural and electronic environment of each metal is mutually influenced by the redox activity of the other.

For the highest Co substitution composition, $LiCoMnO_4$, the cell was charged directly to 5.3 V. Given the full Co substitution, Mn is expected to remain in the +4 oxidation state throughout cycling; however, the Mn K-edge XANES spectra (**Figure S11 c, top**) reveal discernible spectral evolution across the Pristine, Intermediate, and Final Charged species, indicating that the Mn local environment is not entirely static. The corresponding Mn K-edge concentration profile (**Figure S12 c, top**) shows a rapid consumption of the Pristine species at the onset of charge, with the Intermediate species emerging during the lower voltage region, attributed to changes in the Mn redox environment driven by Mn oxidation. The Final Charged species appears toward the end of charge, consistent with further modifications to the Mn local environment induced by Co oxidation processes occurring at higher voltages.

At the Co K-edge (**Figure S12 c, bottom**), the Intermediate species emerges during mid-charge, likely reflecting the indirect influence of Mn oxidation on the local Co environment, while the Final Charged species arising from direct Co oxidation reaches its maximum concentration at the end of charge. Upon discharge, the process exhibits a notably asymmetric and irreversible character: the system does not return to a single-phase Pristine state, and instead a mixture of all three species persists at the end of discharge, with the Pristine species representing the largest fraction. This irreversibility upon discharge is similarly reflected in the Mn K-edge concentration profiles, suggesting that the structural and electronic changes induced during high-voltage charge are only partially reversible in this composition.

## 2.2. S2.2 EXAFS fitting and TM bond length analysis

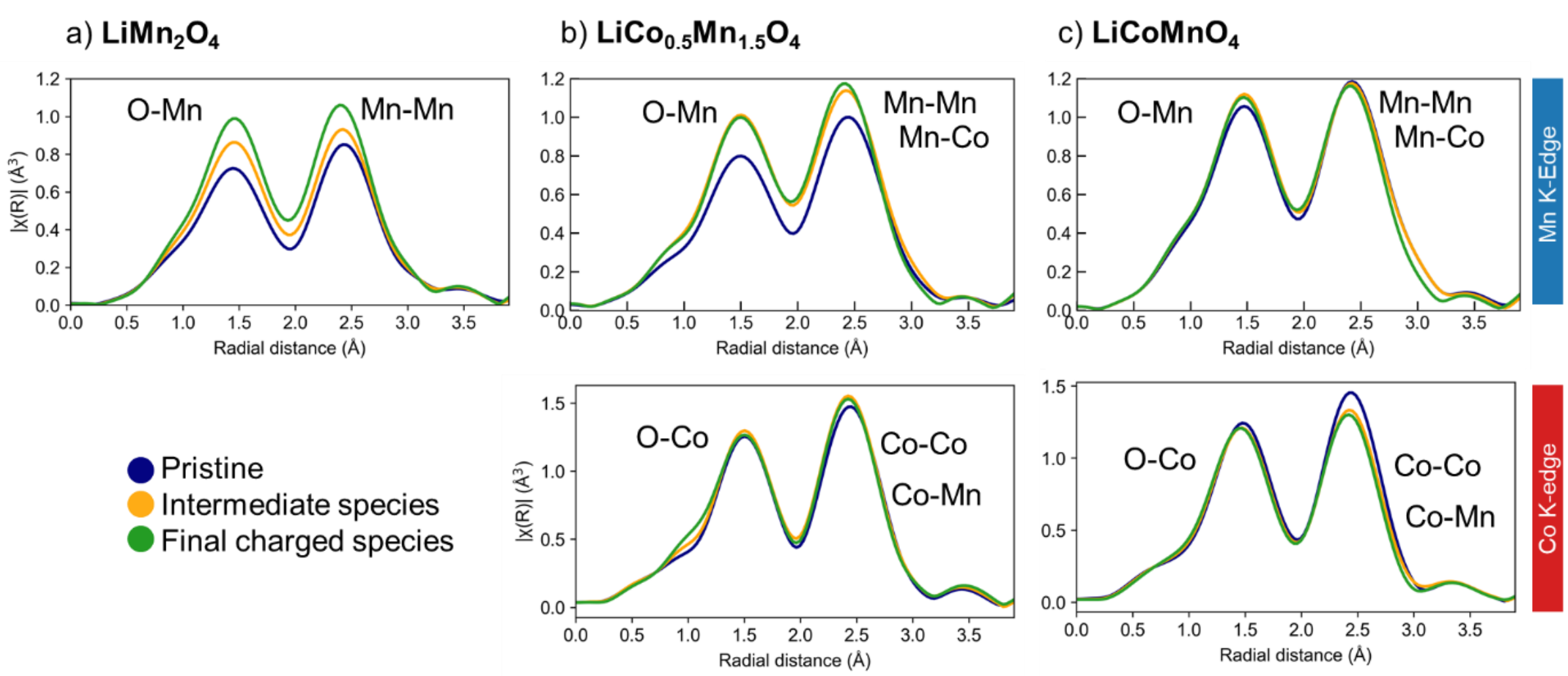

**Figure S13** | Magnitude of the Fourier transforms (FT) of the EXAFS oscillations extracted at Mn (top) and Co (bottom) K-edges of the reconstructed principal components (pristine in blue, intermediate in yellow, and final charged species in green) obtained by MCR-ALS analysis for a) $LiMn_2O_4$, b) $LiCo_{0.5}Mn_{1.5}O_4$, and c) $LiCoMnO_4$. The FTs are not corrected by the phase shifts.

As discussed in the main text (**Section 2.4.2**), only minor changes in the TM-O bond length are observed for all materials, accompanied by a small contraction of the TM-TM bond upon Li extraction and a subsequent expansion upon Li insertion. The fitted bond-length values are summarized in **Tables S2 - S4** for $LiMn_2O_4$, $LiCo_{0.5}Mn_{1.5}O_4$, and $LiCoMnO_4$, respectively. The magnitude of these changes is limited to the resolution of the EXAFS measurements. Therefore, resolving more subtle variations in the TM-O and TM-TM bond lengths would require higher-resolution measurements and more precise analysis.

For $LiMn_2O_4$, the local structure becomes progressively more ordered with increasing state of charge, as evidenced by a systematic increase in peak intensity in the second coordination shell (**Figure S13 a**). This trend is not observed in the other compositions, which exhibit broadly similar peak intensities across the different principal components. An exception is noted at the Mn K-edge of $LiCo_{0.5}Mn_{1.5}O_4$, where a moderate degree of structural ordering is also observed from the Pristine to the Final Charged species, though less pronounced than in $LiMn_2O_4$. This behaviour is further supported by the Debye-Waller factor values extracted for all samples (**Tables S2 – S4**), where $LiMn_2O_4$ exhibits a slight but consistent decrease from the Pristine to the Final Charged species, indicative of reduced structural disorder and progressive ordering upon charge. In contrast, for the Co-substituted compositions $LiCo_{0.5}Mn_{1.5}O_4$ and $LiCoMnO_4$, no significant changes in the Debye-Waller factors are observed across either the TM-O or TM-TM bonds at different states of charge, suggesting that the introduction of Co into the spinel framework imparts a greater degree of structural rigidity, consistent with the formation of more covalent transition metal-oxygen bonds.

**Table S2** | Results of EXAFS fitting parameter of the reconstructed principal components for $LiMn_2O_4$

| Shell | Reconstructed principal components | *N* | $R_{theo}$ | $R_{fit}$ | $\sigma^2$ |
|---|---|---|---|---|---|
| **Mn-O** | Pristine | 6 | 1.947 | 1.91(5) | 0.009(3) |
| | Intermediate | 6 | 1.947 | 1.90(5) | 0.007(3) |
| | Final Charged | 6 | 1.947 | 1.90(5) | 0.006(3) |
| **Mn-Mn** | Pristine | 6 | 2.907 | 2.91(5) | 0.008(3) |
| | Intermediate | 6 | 2.907 | 2.88(5) | 0.007(3) |
| | Final Charged | 6 | 2.907 | 2.86(5) | 0.006(3) |

**Table S3** | Results of EXAFS fitting parameter of the reconstructed principal components for $LiCoMnO_4$

| Shell | Reconstructed principal components | *N* | $R_{theo}$ | $R_{fit}$ | $\sigma^2$ |
|---|---|---|---|---|---|
| **Mn-O** | Pristine | 6 | 1.908 | 1.90(5) | 0.005(3) |
| | Intermediate | 6 | 1.908 | 1.90(5) | 0.004(3) |
| | Final Charged | 6 | 1.908 | 1.90(5) | 0.005(3) |
| **Co-O** | Pristine | 6 | 1.908 | 1.91(5) | 0.004(5) |
| | Intermediate | 6 | 1.908 | 1.89(5) | 0.004(5) |
| | Final Charged | 6 | 1.908 | 1.88(5) | 0.004(5) |
| **Mn-Mn** | Pristine | 3 | 2.849 | 2.91(5) | 0.004(3) |
| | Intermediate | 3 | 2.849 | 2.91(5) | 0.003(3) |
| | Final Charged | 3 | 2.849 | 2.90(5) | 0.003(3) |
| **Co-Co** | Pristine | 3 | 2.849 | 2.88(5) | 0.003(5) |
| | Intermediate | 3 | 2.849 | 2.87(5) | 0.004(5) |
| | Final Charged | 3 | 2.849 | 2.86(5) | 0.004(5) |
| **Mn-Co** | Pristine | 3 | 2.849 | 2.82(5) | 0.004(3) |
| | Intermediate | 3 | 2.849 | 2.80(5) | 0.003(3) |
| | Final Charged | 3 | 2.849 | 2.79(5) | 0.003(3) |
| **Co-Mn** | Pristine | 3 | 2.849 | 2.82(5) | 0.003(5) |
| | Intermediate | 3 | 2.849 | 2.81(5) | 0.004(5) |
| | Final Charged | 3 | 2.849 | 2.79(5) | 0.004(5) |

Table S4 | Results of EXAFS fitting parameter of the reconstructed principal components for $LiCo_{0.5}Mn_{1.5}O_4$

| **Shell** | **Reconstructed principal components** | ***N*** | $R_{theo}$ | $R_{fit}$ | $\sigma^2$ |
|---|---|---|---|---|---|
| **Mn-O** | Pristine | 6 | 1.925 | 1.91(5) | 0.007(3) |
| | Intermediate | 6 | 1.925 | 1.90(5) | 0.005(3) |
| | Final Charged | 6 | 1.925 | 1.90(5) | 0.005(3) |
| **Co-O** | Pristine | 6 | 1.925 | 1.91(5) | 0.002(3) |
| | Intermediate | 6 | 1.925 | 1.90(5) | 0.003(3) |
| | Final Charged | 6 | 1.025 | 1.88(5) | 0.003(3) |
| **Mn-Mn** | Pristine | 4 | 2.874 | 2.91(5) | 0.003(3) |
| | Intermediate | 4 | 2.874 | 2.88(5) | 0.003(3) |
| | Final Charged | 4 | 2.874 | 2.87(5) | 0.003(3) |
| **Co-Co** | Pristine | 4 | 2.874 | 2.85(5) | 0.003(3) |
| | Intermediate | 4 | 2.874 | 2.81(5) | 0.002(3) |
| | Final Charged | 4 | 2.874 | 2.81(5) | 0.002(3) |
| **Mn-Co** | Pristine | 2 | 2.874 | 2.79(5) | 0.003(3) |
| | Intermediate | 2 | 2.874 | 2.78(5) | 0.003(3) |
| | Final Charged | 2 | 2.874 | 2.78(5) | 0.003(3) |
| **Co-Mn** | Pristine | 2 | 2.874 | 2.89(5) | 0.003(3) |
| | Intermediate | 2 | 2.874 | 2.88(5) | 0.002(3) |
| | Final Charged | 2 | 2.874 | 2.88(5) | 0.002(3) |